\documentclass[twocolumn]{aastex61}
\pdfoutput=1 
\usepackage{amsmath,amstext}
\usepackage[T1]{fontenc}
\usepackage{apjfonts} 
\usepackage[figure,figure*]{hypcap}
\usepackage{gensymb}
\usepackage{multirow}
\usepackage{longtable}
\usepackage{xcolor}

\def\arcmin{\hbox{$^\prime$}}
\def\arcsec{\hbox{$^{\prime\prime}$}}

\def\tractor{{\sc Tractor}}

\shorttitle{A spectral database of quasar pairs}
\shortauthors{J~R.~Findlay et al.}

\begin{document}

\defcitealias{2006ApJ...651...61H}{QPQ1}
\defcitealias{2007ApJ...655..735H}{QPQ2}
\defcitealias{2009ApJ...690.1558P}{QPQ3}
\defcitealias{2013ApJ...766...58H}{QPQ4}
\defcitealias{2013ApJ...762L..19P}{QPQ5}
\defcitealias{2013ApJ...776..136P}{QPQ6}
\defcitealias{2014ApJ...796..140P}{QPQ7}
\defcitealias{2016ApJS..226...25L}{QPQ8}
\defcitealias{2017arXiv170503476L}{QPQ9}
\title{Quasars probing quasars X: The quasar pair spectral database}

\author{Joseph R. Findlay}
\affiliation{
University of Wyoming,
Physics \& Astronomy
1000 E. University, Dept 3905
Laramie, WY 82071}
\author{ J. Xavier Prochaska}
\altaffiliation{
University of California Observatories, 
Lick Observatory, 
1156 High Street, 
Santa Cruz, 
California 95064, USA}
\affiliation{
Department of Astronomy and Astrophysics, 
University of California, 
1156 High Street, 
Santa Cruz, 
California 95064, USA}
\author{Joseph F. Hennawi}
\altaffiliation{
Department of Physics, 
Broida Hall, 
University of California, 
Santa Barbara, CA 93106-9530, USA}
\affiliation{
Max-Planck-Institut f{\"u}r Astronomie, 
K{\"o}nigstuhl 17, D-69117 Heidelberg, Germany}
\author{Michele Fumagalli}
\altaffiliation{
Centre for Extragalactic Astronomy,
Department of Physics,
Durham University, 
South Road, 
Durham, 
DH1 3LE, UK
}
\affiliation{
Institute for Computational Cosmology, 
Durham University, 
South Road, 
Durham, DH1 3LE, UK
}
\author{Adam D. Myers}
\affiliation{
University of Wyoming,
Physics \& Astronomy
1000 E. University, Dept 3905
Laramie, WY 82071}
\author{Stephanie Bartle}
\affiliation{
Centre for Extragalactic Astronomy,
Department of Physics,
Durham University, 
South Road, 
Durham, 
DH1 3LE, UK
}
\author{Ben Chehade}
\affiliation{
Centre for Extragalactic Astronomy,
Department of Physics,
Durham University, 
South Road, 
Durham, 
DH1 3LE, UK
}
\author{Michael A. DiPompeo}
\affiliation{
Department of Physics and Astronomy,
Dartmouth College, 
6127 Wilder Laboratory, 
Hanover, NH 03755, USA
}
\author{Tom Shanks}
\affiliation{
Centre for Extragalactic Astronomy,
Department of Physics,
Durham University, 
South Road, 
Durham, 
DH1 3LE, UK
}
\author{Marie Wingyee Lau}
\affiliation{
Department of Astronomy and Astrophysics, 
UCO/Lick Observatory, University of California, 
1156 High Street, Santa Cruz, CA 95064 
}
\author{Kate H. R. Rubin}
\affiliation{
San Diego State University, 
Department of Astronomy, 
San Diego, CA 92182, USA
}
\begin{abstract}
The rare close projection of two quasars on the sky provides the opportunity to study the host 
galaxy environment of a foreground quasar in absorption against the continuum emission of a 
background quasar. For over a decade the "Quasars probing quasars" series has utilized this 
technique to further the understanding of galaxy formation and  evolution in the presence of a 
quasar at $z>2$, resolving scales as small as a galactic disc and from bound gas in the 
circumgalactic medium to the diffuse environs of intergalactic  space. Presented here, is the public 
release of the quasar pair spectral database utilized in these studies. In addition to projected 
pairs at $z>2$, the database also includes quasar pair members at $z<2$, gravitational lens 
candidates and quasars closely separated in redshift that are useful for small-scale clustering 
studies. In total the database catalogs 5627 distinct objects, with 4083 lying within 5\arcmin of at 
least one other source. A spectral library contains 3582 optical and near-infrared spectra for 3028 
of the cataloged sources. As well as reporting on 54 newly discovered quasar pairs, we outline the 
key contributions made by this series over the last ten years, summarize the imaging and 
spectroscopic data used for target selection, discuss the target selection methodologies, describe 
the database content and explore some avenues for future work. Full documentation for spectral 
database, including download instructions are supplied at 
\url{http://specdb.readthedocs.io/en/latest/}
\end{abstract}

\keywords{catalogs --- surveys --- galaxies: evolution --- galaxies: halos --- quasars: absorption 
lines --- quasars: general}

\section{Introduction}
An understanding of the processes by which galaxies accrete, expel and recycle gas is as essential 
to a complete theory of galaxy evolution as population demographics and star formation. At its 
largest scales, the Universe is seen as a living network of filaments blossoming with clusters and 
galaxies at their intersections. This "Cosmic web" contains 90 percent of all the baryonic material 
in the Universe and supplies the fuel for galaxy formation along its filaments. The journey of this 
gas into the interstellar medium (ISM) occurs through the circumgalactic medium (CGM), a 
gravitationally bound gas reservoir, distinct from the ISM but contained within a region similar in 
extent to a galaxy's own virial radius. In their recent review, \citet{2017ARA&A..55..389T} liken 
the CGM to a galactic utility provider, acting as the galactic fuel tank, waste dump and recycling 
center simultaneously.

The CGM is diffuse and therefore difficult to detect in emission, and has traditionally received 
less attention than the ISM and the intergalactic medium (IGM). However, interest in the CGM has 
undergone somewhat of a revolution in recent years. Galactic gas flows, which by necessity must pass 
through the CGM, appear to be at the heart of a myriad of unresolved and compelling issues including 
the ``missing baryon'' problem \citep[e.g.][]{2017ApJS..230....6K}, the metal census 
\citep[e.g.][]{2014ApJ...786...54P}, the quenching of star formation in passive galaxies and the 
perpetuation of star formation in star forming galaxies 
\citep[the so called red-blue dichotomy e.g.][]{2011ApJ...743...10B,2016ApJ...833..259B}. 

Advances in the modeling of galactic flows has prompted large strides forward in the observational 
capabilities to detect them, notably with the {\it Cosmic Origins Spectrograph} \citep[COS;][]
{2013ApJ...777...59T,2013ApJS..204...17W} on the {\it Hubble Space Telescope} (HST). Evidence for 
accretion flows comes from line widths in HI and various other species, which reveal velocity 
dispersions in halo gas that show it to be bound 
\citep[e.g.][]{2013ApJ...777...59T,2014ApJ...796..136B,2017ApJ...835..267H}. In the Milky Way, 
unmistakable evidence for accretion comes in the form of blueshifted high velocity clouds (HVCs) 
\citep[e.g.][]{2004ApJS..150..387S,2011Sci...334..955L}. In other galaxies the evidence has been 
more difficult to ascertain presumably because individual flows contain insufficient mass or are too 
diffuse or both \citep{2012ApJ...747L..26R}. There are a handful of examples in which diffuse gas 
has been detected in emission around other galaxies 
\citep{2014Natur.506...63C,2014ApJ...786..106M,2015Sci...348..779H,2017MNRAS.471.3686F} but for the 
most part the community has relied on absorption line spectroscopy 
\citep{2012ApJ...747L..26R,2016ApJ...820..121B,2017arXiv170501543W}.

On the other hand observations of pristine gas in the IGM are rare \citep{2011Sci...334.1245F} and 
the metalicity of the gas in all phases of the CGM 
\citep{2013ApJ...777...59T,2014ApJ...786...54P,2017ApJ...837..169P} is a sure indication that it has 
at some point passed through the ISM. This evokes a picture in which a significant amount of 
accreted material is recycled, blown out of the ISM and into galactic halos via feedback-triggered 
outflows. The evidence for outflows is large 
\citep{2010ApJ...717..289S,2014ApJ...794..156R,2015ApJ...799L...7F,2017arXiv170501543W}. What is 
less well known is how outflows interact with the CGM, and beyond, to transport processed matter and 
energy, and to what extent these processed materials are recycled in subsequent star formation.

The extensive progress of this field in recent years is examined comprehensively in a number of 
recent reviews \citep{2012ARA&A..50..491P,2017ARA&A..55..389T,2017ASSL..430.....F}. However a 
complete picture of the CGM in the context of galaxy evolution is unachievable without broadening 
the discussion towards the extreme and evanescent phases of galaxy evolution such as nuclear 
starbursts, mergers and active galaxies or quasars. The latter has been the subject of a series of 
nine papers known as the {\it Quasars Probing Quasars} (QPQ) project 
\citep[][henceforth QPQ1-QPQ9]{2006ApJ...651...61H,2007ApJ...655..735H,2009ApJ...690.1558P,
2013ApJ...766...58H,2013ApJ...762L..19P,2013ApJ...776..136P,2014ApJ...796..140P,2016ApJS..226...25L,
2017arXiv170503476L}. Just as the Cosmic web or a galactic halo can be observed in absorption 
against the continuum emission of a distant quasar, so too can the CGM of a quasar host when a 
background quasar is projected close to the line of sight of a foreground quasar. This is in fact 
the idea behind the QPQ project, which aims to elucidate the CGM in the massive 
$M\approx 10^{12.5}\,M_{\odot}$ halos that host quasars at $z\gtrsim 2$
\citep[e.g.][]{2015MNRAS.453.2779E}.

A reoccurring conclusion throughout the QPQ series is that a quasar's ionizing continuum illuminates 
surrounding gas in an anisotropic fashion. Excess absorbers are found transverse to quasar sight 
lines \citepalias{2006ApJ...651...61H} and the quasar-absorber clustering signal measured in 
transverse directions has been shown to over predict the number of absorbers in line of sight 
directions by several times \citepalias{2007ApJ...655..735H}. This result holds to $\mathrm{Mpc}$ 
scales \citepalias{2013ApJ...776..136P} and beyond 
\citep[][Sorini et al.\ 2018, in prep.]{2013JCAP...05..018F}. Furthermore, if this cool gas were to 
be illuminated by the quasar radiation field then it should emit Ly$\alpha$ photons, either from 
fluorescent recombinations, resonant scattering, or by Ly$\alpha$ cooling radiation. This has only 
rarely been found on large scales 
(\citetalias{2013ApJ...766...58H};\,\citealt{2014Natur.506...63C,2014ApJ...786..106M,
2015Sci...348..779H}). Other lines of inquiry have failed to detect the transverse proximity effect 
in background quasar spectra despite numerous attempts \citep{1995MNRAS.277..235F,
2001MNRAS.328..653L,2004ApJ...610..105S,2004ApJ...610..642C,2013ApJ...762L..19P}. 

Large covering fractions of HI extend to impact parameters of $R_{\perp}\lesssim 200\,\mathrm{kpc}$ 
likely coinciding with the virial radii of halos \citepalias{2013ApJ...762L..19P,
2013ApJ...776..136P,2016ApJS..226...25L,2014ApJ...796..140P}. This gas represents up to a third of 
the total baryonic mass within the CGM, approaching the mass of baryons in the ISM 
\citepalias{2014ApJ...796..140P}. Beyond $R_{\perp}\gtrsim 500\,\mathrm{kpc}$ the mass of cool gas 
remains substantial. If it is possible to generalize these findings to all coeval massive galaxies, 
then a significant amount of all optically thick gas is found within the extended regions around 
massive galaxies at $z \approx 2$ \citepalias{2013ApJ...776..136P}. Indeed, this phenomenon has no 
obvious affinity with coeval galaxy populations in other mass regimes 
\citepalias{2014ApJ...796..140P} and must surely indicate a divergence of evolutionary processes.
 
Clues to the provenance of this cool halo gas can be explored via its metal content. Incidences of 
low-ion metal absorption systems follow closely with the results of HI above, i.e.\ the gas is 
enriched out to impact parameters of at least $\mathrm{1\,Mpc}$ and shows the strongest signatures 
of metal enrichment at $R_{\perp}\lesssim 200\,\mathrm{kpc}$ 
\citepalias{2013ApJ...762L..19P,2013ApJ...776..136P,2016ApJS..226...25L,2014ApJ...796..140P}. 
Stacked spectra reveal average profiles of both low and high ions with systematically larger 
equivalent widths than any other known galaxy population \citepalias{2014ApJ...796..140P}. Strong 
signatures of $\alpha$-elements implicate core-collapse supernovae as the progenitors of this gas 
and point to star formation histories similar to massive ellipticals, which are thought to be 
the modern-day descendants of quasars \citepalias{2016ApJS..226...25L,2017arXiv170503476L}. This may 
link the enrichment of the CGM in quasar hosts at least in part to their own ISM but poses further 
questions regarding transport mechanisms.

Both quasar and star formation feedback are invoked as the transport mechanisms required to move gas 
out of the ISM and into the CGM. However QPQ finds no obvious evidence for any single dominant 
process. While kinematics are extreme and in some cases suggestive of violent outflow 
\citepalias{2009ApJ...690.1558P,2016ApJS..226...25L}, on the whole there is no need to appeal to 
anything beyond the gravitationally supported dynamics expected of massive halos 
\citepalias{2014ApJ...796..140P,2017arXiv170503476L}. Simple arguments concerning timescales and 
energetics show that episodes of quasar activity are insufficient to place metals at 
$\sim \mathrm{Mpc}$ distances within a single duty cycle
\citepalias{2009ApJ...690.1558P,2014ApJ...796..140P}. Furthermore cool gas is present, albeit in 
lower quantities, in the halos of quiescent galaxy populations so ongoing quasar activity is not a 
prerequisite for its existence. On the other hand the average gas covering fraction may be 
correlated with quasar bolometric luminosity \citep{2015MNRAS.452.2553J}, making it difficult to 
argue against the connection between cool halo gas and quasar activity. Far from ruling out quasar 
feedback as a transport mechanism completely, it is more appealing to invoke it as an intermittent 
agent operating on $\sim 100\,\mathrm{kpc}$ scales.

Outflows supported by star formation feedback are found in a variety galaxy populations across time 
\citep{1998ApJ...508..539P,2005ApJ...621..227M,2009ApJ...692..187W,2014ApJ...794..156R}. Ongoing 
star formation however, does not appear to be precondition of a cool enriched CGM. Early type 
galaxies contain a large reservoir of cool enriched halo gas \citep{2012ApJ...758L..41T} and there
is scant evidence to link the strength or prevalence of optically-thick absorbers to star formation 
rate \citepalias{2014ApJ...796..140P}. Indeed $z\sim 2$ quasar hosts do not show evidence for 
increased star formation \citep{2013A&A...560A..72R,2017MNRAS.472.2221S}. Furthermore the large 
impact parameters to which cool gas extends puts it beyond the influence of any single star 
formation episode. The presence of heavy elements in the CGM serves as strong evidence that it has 
at some point passed through the ISM, but as with quasar feedback this is likely to happen over 
integrated episodes limited to $\sim 100\,\mathrm{kpc}$ scales. 

The presence of heavy elements in the cool halo gas on $\mathrm{Mpc}$ scales means that it cannot be 
entirely explained by accretion of pristine gas from the cosmic web. Rather it may arise from 
galactic winds driven by star formation in low-mass satellite galaxies that deposit their metals in 
cosmic filaments. Modeling of these processes has begun to be able to match the covering fractions 
of cool gas seen in observations \citep{2016MNRAS.461L..32F}, which alleviates the need for feedback 
processes in the host galaxy to act at these distances.

The QPQ series has contributed to a boom in the understanding of the CGM in the massive galaxies 
that host quasars at $z>2$. To facilitate this work the project has put together a database of over 
2000 projected quasar pairs with transverse separations at foreground quasar redshifts ranging from 
between tens of physical kpc to a few Mpc. Equivalently the database probes scales as small as a 
galactic disc, to unbound gas in the surrounding IGM, to everything in between. The QPQ 
collaboration here presents its extensive database of quasar pairs and their spectra. This 
submission adopts the $\Lambda$CDM cosmology of the \citet{2016A&A...594A..13P} and reports all 
magnitudes on the AB system.  Section~\ref{sec:experiment} recalls the basic principles behind the 
QPQ experiment, section~\ref{sec:specselect} discusses the selection of quasar pairs in 
spectroscopic redshift surveys, section~\ref{sec:imagingSelection} discusses quasar pair selection 
in imaging data, section~\ref{sec:atlaswise} presents recent results from the latest QPQ quasar pair 
search using ATLAS, SDSS and WISE imaging, section~\ref{sec:database} describes the content and 
architecture of the quasar pair catalog and accompanying spectral database and 
section~\ref{sec:summary} provides a summary of this submission and lays out some avenues for the 
future exploitation of the QPQ database.

\section{The Experiment}\label{sec:experiment}

\begin{figure}
   \centering
   \includegraphics[scale=0.35]{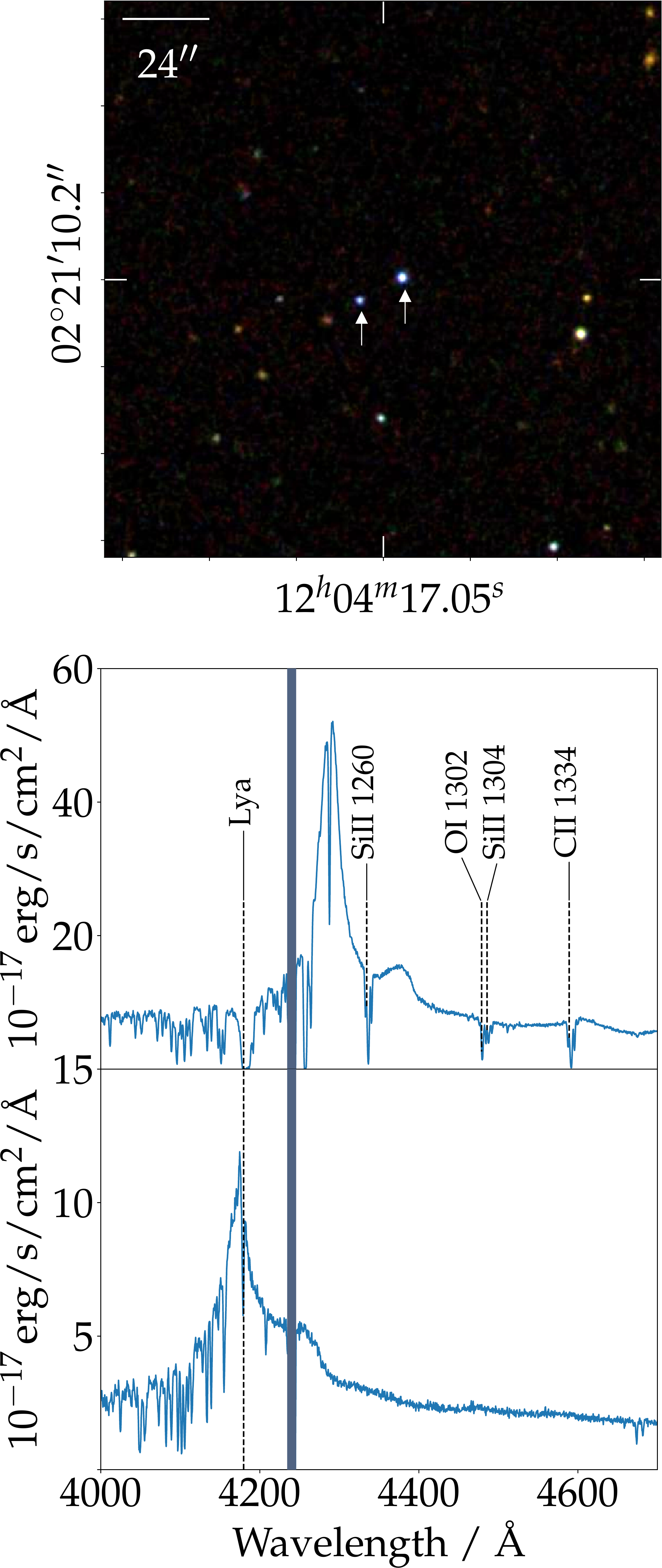}
   \caption{Illustrating the basic observation behind the QPQ experiment. The upper panel 
   shows the projected quasar pair J1204+0221 at redshifts of z=$2.44$ and $z=2.53$. 	 
   Photons from the background quasar pass through the halo of the foreground quasar and 
   ionize HI and other atomic gas on their path. In the lower panel this process in 
   manifested in the form of absorption line transitions in the spectrum of the 
   background quasar (upper) at transition wavelengths corresponding to the redshift of 
   the foreground quasar (lower). The shaded areas mask regions of bad
   data.}\label{fig:example}
\end{figure}

To begin, it is useful to briefly review the basic principle behind the QPQ experiment. The upper 
panel of Figure~\ref{fig:example} shows the projected quasar pair J1204+0221 in SDSS 
\cite[Sloan Digital Sky Survey;][]{2000AJ....120.1579Y} imaging. The pair members are separated 
by an angular distance of $\theta\simeq 13.28^{\prime\prime}$ on the sky and are located at 
foreground and background redshifts of $z=2.44$ and $z=2.53$ respectively. From the adopted 
cosmology it follows that the sightline of the background quasar passes the foreground quasar at a 
transverse distance of $R_{\perp}\simeq 110\,\mathrm{pkpc}$. On their way to the observer, 
background quasar photons are absorbed and scattered by gas associated with the foreground quasar's 
halo and may ionize HI and other atomic species. The absorption features are seen as electronic 
transitions, primarily at far-ultra-violet (UV) rest-frame energies, in the spectrum of the 
background quasar at transition wavelengths corresponding to the redshift of the absorbing gas. This 
is shown in the lower panel, where Ly$\alpha$ absorption is clearly seen in the background spectrum 
at the redshift of the foreground quasar. This is accompanied by a series of labeled metal 
transitions.

A major goal of the QPQ project has been to search for projected quasar pairs in optical 
spectroscopic and photometric surveys. The blue cutoff in the optical $u$ band falls at 
$\sim3500$\AA, which corresponds to the wavelength of the redshifted Ly$\alpha$ transition at 
$z\sim 2$. This is the limit at which one can reasonably detect atomic hydrogen at optical 
wavelengths and in large, ground-based surveys such as the SDSS. QPQ pair searches have therefore 
been focused towards pairs with foreground members within this redshift regime. Inevitably however, 
pairs outside of this redshift regime have found their way into the QPQ sample as false positives. 
These pairs are also included in the database, as are close physical binaries and gravitational lens 
candidates.

\section{Quasar pair selection from spectroscopic surveys}\label{sec:specselect}

A projected quasar pair is confirmed once spectra have been obtained that unambiguously reveal the 
spectral type of each pair member. In an age where spectroscopic redshift surveys have become a 
mainstay in extragalactic exploration, numbers of known quasar pairs have seen a corresponding spurt 
in growth. 

Over the years, public spectroscopic survey projects have observed and cataloged over half a million 
quasars to redshifts in excess of $z=7$. Unquestionably, the largest compendium of quasar 
spectroscopy has been provided by the SDSS. Using straightforward spatial matching, QPQ has relied 
heavily on the annual spectroscopic data releases of the SDSS to select quasar pairs at separations 
of $\gtrsim$1\arcmin. Due to fiber collisions, some pairs at closer separations require alternate 
techniques, which are discussed later in this section.

\subsection{The SDSS Legacy Surveys}

The original SDSS \citep{2000AJ....120.1579Y} ran between the years 2000-2008 and provides both 
imaging and spectroscopic coverage over $\sim 12,000\,\mathrm{deg^2}$ of the sky. The two components 
of the survey, SDSS-I and SDSS-II, ran consecutively on a dedicated 2.5\,m telescope located at the 
Apache Point Observatory, in New Mexico \citep{2006AJ....131.2332G}. Collectively, these surveys are 
now known as the SDSS Legacy survey (SDSS-LS).

SDSS-I ran during the first five years of operation and carried out multicolor $u\ 
(3551\,\text{\AA})$, $g\ (4686\,\text{\AA})$, $r\ (6166\,\text{\AA})$, $i\ (7480\,\text{\AA})$, $z\ 
(8932\,\text{\AA})$ \citep{1996AJ....111.1748F,2002AJ....123..485S,2010AJ....139.1628D} imaging and 
targeted spectroscopic follow-up within the imaging footprint. SDSS-II extended SDSS-I imaging 
footprint towards the Galactic plane.

The imaging and astrometric pipelines are described by \cite{2001ASPC..238..269L} and 
\cite{2003AJ....125.1559P} respectively. Photometric calibration is tied to a standard star network 
\citep{2001AJ....122.2129H,2002AJ....123.2121S} and is refined to the $\sim 1$ per cent level via a 
global re-calibration \citep{2008ApJ...674.1217P}. The $95\%$ completeness point source imaging 
depths are $u=22.0$, $g=22.2$, $r=22.2$, $i=21.3$ and $z=20.5$. 

\subsection{The SDSS Baryon Oscillation Spectroscopic Survey}

The SDSS-LS was followed by SDSS-III, incorporating the Baryon Oscillation Spectroscopic Survey 
\citep[BOSS;][]{2013AJ....145...10D}, which was designed to measure the scale of baryon acoustic 
oscillations (BAO) using quasars to trace the clustering of mass. SDSS-III extended the multi color 
imaging of the SDSS-LS by a further $\sim 2500\,\mathrm{deg}^2$ with the same telescope and camera 
between the years 2008-2009. All the SDSS imaging data was uniformly reduced with an improved sky 
subtraction and released under DR8 \citep{2011ApJS..193...29A}, bringing the total imaging footprint 
to $14,555\, \mathrm{deg^2}$. 

The remainder of the survey (2009-2014) was devoted to spectroscopy. The SDSS-LS spectrographs were 
upgraded \citep{2013AJ....146...32S} with new higher-efficiency volume holographic gratings, 
fully-depleted red CCDs with superior red response, blue CCDs with improved blue response and 1000 
new fibers. Continuous wavelength coverage is provided between $3600$-$10,400\,\text{\AA}$ at 
resolutions between 1560-2270 in the blue channel and 1850-2650 in the red channel. Improved 
photometric selection algorithms were used to target 1.5 million luminous galaxies 
\citep{2016MNRAS.455.1553R} and 150,000 new quasars in spectroscopy 
\citep{2012ApJS..199....3R,2017A&A...597A..79P}. 

\subsection{The 2dF QSO Redshift Survey}

Running in partial overlap with the SDSS surveys was the 2 degree Field (2dF) QSO Redshift Survey 
\citep[2QZ;][]{2004MNRAS.349.1397C}. The QPQ search for quasar pairs in 2QZ was a similar search for 
spatially coincident catalog positions.

2QZ provides a homogeneous quasar catalog flux-limited to $16.00<b_J<20.85$. Candidates were color-
selected via multi-wavelength $u$, $b_J$, $r$ photometry from automated plate measurement (APM) of 
UK Schmidt Telescope (UKST) photographic plates. These candidates were then observed by the 2dF 
instrument, a multi-object spectrograph at the Anglo-Australian Telescope (AAT). The 2QZ catalog 
comprises of 23,338 quasars spanning a redshift range of $0.3\lesssim z\lesssim 3.0$. The footprint 
covers a total area of $721.6\, \mathrm{deg}^2$ and is arranged over two contiguous strips of 
$75\degree \times 5\degree$ in area across the southern Galactic cap, centered on 
$\delta=-30\degree$ and northern Galactic cap centered on $\delta=0\degree$. 

\section{Quasar pair selection in imaging surveys}\label{sec:imagingSelection}

The 2QZ and SDSS surveys have made searching for quasar pairs a trivial endeavor with, however, the 
caveat that this method is inherently biased against close pairs due to the finite distance at which 
pairs of spectroscopic fibers can be placed from one another. 

Approximately $30\%$ of the SDSS-LS and BOSS are covered spectroscopically in more than one epoch. 
These regions do not suffer from the effects of {\it fiber collisions} and close quasar pairs in 
these footprints can be drawn directly from the spectroscopic catalogs. Additionally, the 2QZ NGC 
region shares approximately half of its footprint with SDSS-LS, allowing 2QZ quasars to be assigned 
SDSS quasar pairs. 

For the most part, however, fiber collisions prevent objects with separations of $<55\arcsec$, 
$<62\arcsec$  and $<30\arcsec$ from being observed simultaneously in SDSS-LS, BOSS and 2QZ 
respectively \citep{2003AJ....125.2276B,2013AJ....145...10D,2002MNRAS.333..279L}. Therefore in most 
cases, close quasar pairs are selected by spatially matching to photometrically targeted objects 
with quasar-like colors in close proximity to spectroscopically cataloged quasars. Photometric 
candidates are later followed-up spectroscopically to confirm the pair.

QPQ has used SDSS imaging to select photometric quasar pair members in this way via three distinct 
methods. \citet{2006AJ....131....1H} described a means to select pairs at similar redshifts via a 
$\chi^2$ selection statistic, which utilizes the fact that the rest-frame UV to optical spectral 
energy distributions of quasars follow a remarkably tight color redshift relation. If one neglects 
the relatively small intrinsic scatter of the population about this relation, then the fluxes of a 
pair of quasars at the same redshift should be related by a single proportionality constant across 
all bands. Variation in this proportionality can then be approximately attributed to observational 
error. One can find the constant of proportionality that minimizes the sum of the differences in 
flux between pairs across all bands. In this way it is possible to efficiently select close pairs of 
quasars at similar redshifts.

To select close quasar pairs at differing redshifts, \citet{2006AJ....131....1H} 
\citep[see also][]{2007ApJ...658...99M,2008ApJ...678..635M,2017MNRAS.468...77E} made use of the 
various photometric catalogs collated by \citet{2004ApJS..155..257R,2009ApJS..180...67R,
2009AJ....137.3884R,2015ApJS..219...39R}. Richards et.\ al.\ used kernel density estimation to model 
the probability density functions of stars and quasars in the 4-D SDSS color space. The likelihood 
that a given photometric object originates from either of these two distributions is combined with 
prior knowledge of quasar and stellar number densities using Bayes' theorem. The result is a 
probabilistic classification of an object as either a star or a quasar. QPQ mined these catalogs for 
both photometric-photometric and photometric-spectroscopic pairs with large quasar probabilities. 
Candidates were then followed up spectroscopically.

\citet{2011ApJ...729..141B} compiled a targeting catalog of photometric quasar candidates for BOSS 
using a non-parametric Bayesian classifier that represents an advance on the Richards et al.\ 
technique by approximating the underlying density distributions of stars and galaxies in flux space 
via an extreme-deconvolution. The classification code, {\sc XDQSO}, was extended to provide 
probabilistic selection over arbitrary redshift intervals as well as incorporating UV and near-
infrared (IR) information \citep[{\sc XDQSOz};][]{2012ApJ...749...41B}. Further details of the 
continuing effort to discover both photometric-spectroscopic and photometric-photometric quasar 
pairs is detailed in \citet{2004PhDT........23H} and 
\citet{2006AJ....131....1H,2010ApJ...719.1672H}.

\section{New pairs from SDSS, ATLAS \& WISE}\label{sec:atlaswise}

The most recent QPQ search for photometric-photometric pairs was conducted in optical imaging from 
the SDSS and VST ATLAS \citep{2015MNRAS.451.4238S} surveys combined with mid-IR data from the Wide-
Field Infrared Survey Explorer \citep[WISE;][]{2010AJ....140.1868W}. Unlike the searches described 
in previous sections, this search has not been discussed in any previous publication, it is 
described here in detail for the first time.

\subsection{The VST ATLAS Survey}

ATLAS is an optical $u$, $g$, $r$, $i$, $z$ survey carried out by OmegaCam 
\citep{2011Msngr.146....8K} on the European Southern Observatory's (ESO) $2.61\,\mathrm{m}$ VLT 
Survey Telescope \citep[VST;][]{2012SPIE.8444E..1CS} at the Cerro Paranal Observatory in Chile. 
ATLAS has completed its final $4711\,\mathrm{deg}^2$ footprint in all filters over two contiguous 
regions covering the northern and southern Galactic caps during 6 years of observations. The premise 
of the ATLAS survey is to provide imaging in the southern hemisphere with equivalent depth and 
better image quality than SDSS.

The ATLAS data is reduced by end-to-end astrometric and photometric pipelines run by the Cambridge 
Astronomical Survey Unit\footnote{\url{http://casu.ast.cam.ac.uk/}} (CASU) and archived by the Wide 
Field Astronomy Unit (WFAU) in the OmegaCam Science Archive\footnote{\url{http://osa.roe.ac.uk/}} 
(OSA). 

\subsection{The Wide-Field Infrared Survey Explorer}

WISE was launched in December 2009 and from February to August 2010, surveyed the entire sky in four 
mid-IR bands W1 ($3.4\,\mu\mathrm{m}$), W2 ($4.6\,\mu\mathrm{m}$), W3 ($12\,\mu\mathrm{m}$) and W4 \
($22\,\mu\mathrm{m}$). This initial survey release, termed the "AllSky" release, represents a 
significant step forward in exploration of the mid-IR sky at these wavelengths. The 5$\sigma$ AB 
point source sensitivity limits are deeper than 19.1, 18.8, 16.4 and 14.5 mag in W1-W4 respectively, 
providing over a hundred times the sensitivity of the $12\mu\mathrm{m}$ band of the InfraRed 
Astronomical satellite \citep[IRAS;][]{1984ApJ...278L...1N}. 

In September 2010 the cryogen cooling the W3 and W4 instruments depleted, ending the full four band 
mission. The W1 and W2 band missions continued through February 2011 encompassing both the AllWISE 
and Near Earth Object WISE \citep[NEOWISE;][]{2011ApJ...731...53M} data releases. After a period of 
hibernation this was followed by the NEOWISE Reactivation mission
\citep[NEOWISER;][]{2014ApJ...792...30M} in October 2013, which is still in operation at the time of 
writing (March 2018).

\subsection{Data preparation \& candidate selection}

SDSS and ATLAS differ widely to WISE in depth, resolution and wavelength and as a result there are 
significant benefits to forced photometry in WISE images at optical positions compared to 
traditional positional catalog matching. The angular resolution of the WISE imaging is 
diffraction-limited and at long mid-IR wavelengths this translates into a resolving power of several 
arcseconds. Conversely, the angular resolution of the optical imaging catalogs are seeing limited, 
translating to roughly 1.0\arcsec\ at the median. Forced photometry ensures that the optical and 
WISE measurements are linked to a consistent set of sources, whereas catalog-matching would 
inevitably lead to a tail of wide separation erroneous positional matches.

Furthermore the bulk of SDSS and ATLAS quasar candidates lie towards the limiting depths of the 
respective catalogs. Many of these sources are undetected at the shallower limiting depths of the 
AllWISE public release catalogs. Measurements of these faint or even undetected sources in WISE are 
just as scientifically valuable as a significant detection, particularly for the sophisticated 
statistical selection techniques described in section \ref{sec:imagingSelection}.

The full-depth coadded WISE images are available as part of the AllWISE data release. These 
full-depth images are convolved by the point spread function (PSF) during the coaddition process. 
This step is included to improve the detection of isolated point sources but it is inappropriate for 
other applications such as forced photometry, since the blurring of the PSF decreases the available 
signal-to-noise. \citet{2014AJ....147..108L} provide an independent reduction of the AllWISE data 
products, which consists of full-depth coadds at the full instrument resolution. These data 
products, termed the ``unWISE'' coadds, were used to assemble a catalog of mid-IR photometry forced 
at the sites of over 400 million SDSS sources \citep{2016AJ....151...36L}. Closely following this 
release, the {\sc XDQSOz} selection code was updated to incorporate the unWISE imaging and a catalog 
of over 5 million photometric quasar candidates was generated over the SDSS footprint 
\citep{2015MNRAS.452.3124D}. This candidate list served as the starting point for the photometric 
pair search in SDSS and WISE.

More recently \citet{2017AJ....153...38M} have undertaken further reprocessing of AllWISE, 
incorporating three years of NEOWISER imaging into the unWISE framework. The coadded data products 
provide significant gains in depth over the original unWISE release and, in addition, time dependent 
artifacts such as moon contamination are largely eliminated because of the inclusion of multiple 
epochs.

Band merged ATLAS source catalogs provided by CASU were force photometered at the W1 and W2 
bandpasses in the unWISE coadds of \citet{2017AJ....153...38M}. The forced photometry was performed 
by the \tractor\ (Lang et al.\ in preparation). The \tractor\ is an innovative code for inference 
modeling of astronomical sources. The premise is the optimization of a likelihood for the source  
properties given a set of imaging data and an informative noise model. The details of the \tractor\ 
implementation running in forced photometry mode on the unWISE coadds is discussed by 
\citet{2016AJ....151...36L}. The QPQ \tractor\ run follows this closely.

QPQ used the updated {\sc XDQSOz} selection code to construct a catalog of quasar candidates from 
ATLAS and forced unWISE imaging over a fraction of the ATLAS footprint. This catalog, along with the 
SDSS-unWISE catalog described above were then mined for close pairs of objects with $>90\%$ quasar 
probabilities. Candidates were selected to $r<22$ but emphasis was typically placed on selecting 
pairs with $r<21$ and with foreground members at $z>3$. All pairs within 1\arcmin\ were considered 
for follow up, but priority was given to pairs within $<30$\arcsec.

\begin{figure*}
   \centering
   \includegraphics[scale=0.65]{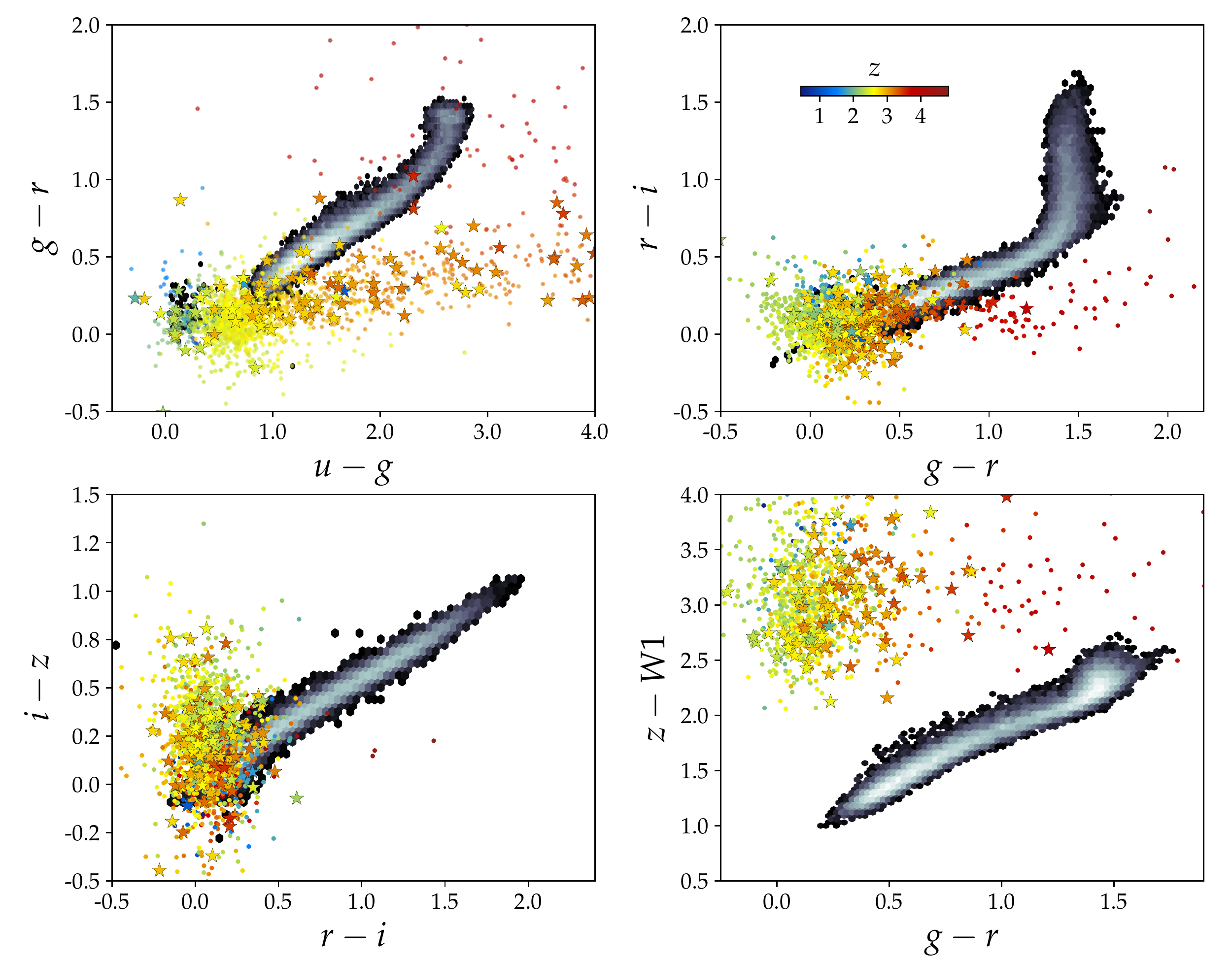}
   \caption{Color-color plots showing the stellar locus in gray scale and the quasar pair 
   database color coded according to redshift. The benefit of introducing the WISE W1 
   band to SDSS $ugriz$ color combinations is seen in the bottom right hand panel, where 
   quasars and stars occupy distinct regions of color-color space. The star shaped 
   symbols show quasar pairs discovered during the recent QPQ search in ATLAS, SDSS and 
   WISE imaging. The circles show the remainder to the QPQ pair database. No attempt
   has been made here to place the ATLAS photometry on to the SDSS system; at the scales
   presented here the differences are completely negligible.}
   \label{fig:colorplots}
\end{figure*}

The benefits of combining WISE mid-IR data into the established $ugriz$ optical quasar selection 
method has been explored since the WISE AllSky data release in March 2012 
\citep[e.g.][]{2012AJ....144...49W}. Among the first projects to explore this practically were the 
SDSS-IV Extended Baryon Oscillation Spectroscopic Survey \citep[eBOSS;][]{2015ApJS..221...27M} and 
the 2dF Quasar Dark Energy Survey pilot \citep[2QDESp;][]{2016MNRAS.459.1179C}, a spectroscopic 
survey of quasar candidates selected in ATLAS and WISE imaging. These projects showed the utility of 
the infrared excess in quasar SEDs to achieve large separations between the quasar and stellar loci. 
The utility of the WISE mid-IR imaging can be seen in Figure~\ref{fig:colorplots}, where the quasar 
pair database is plotted along side stars for four different color-color combinations including the 
SDSS and WISE W1 passbands. Only where WISE data is incorporated, as in the lower right-hand panel, 
is there a significant distinction between the stellar and quasar loci.

\subsection{Spectroscopic follow up}

Candidates at declination $>-20~\rm deg$ (all those presented here) were observed on the 
Intermediate dispersion Spectrograph and Imaging System \citep[ISIS;][]{1990SPIE.1235..790J}, 
mounted at the Cassegrain focus of the 4.2\,m William Herschel Telescope (WHT). The observing run 
was conducted on 22 nights between December 2015 and March 2016 as part of program 2015/P1 and 
2016/P6. Approximately three nights were lost due to poor weather. Usable nights were for the most 
part clear but not always photometric.

The instrument was configured with R300B and R316R gratings centered at 4230\,\AA\ and 6400\,\AA\ in 
the blue and red arms respectively, with the D5300 dichroic filter in place over both channels. A 
1\arcsec slit width combined with a factor 2 binning in the dispersion direction at readout time, 
resulted in continuous wavelength coverage over $3000-7630$\,\AA\ with average resolutions of 
$R\sim 4.1$\AA\ and $R\sim 3.8$\AA\ in the blue and red channels respectively.

The data were reduced using the Low-Redux  
pipeline\footnote{\url{http://www.ucolick.org/~xavier/LowRedux/}} extended to work with the ISIS 
detectors. This pipeline performs basic calibrations (bias, flat-fielding, and wavelength 
calibration), and extracts flux-calibrated 1D spectra which are then coadded across multiple 
exposures. All spectra were visually inspected to separate stars from quasars, and to measure the 
quasar redshifts by superposing a quasar template to the data. 

In total, 69 photometrically selected pairs in ATLAS were observed on WHT, resulting in 15 new 
projected quasar pairs. The remaining observing time was devoted to observations of candidates 
selected from SDSS, yielding the discovery of 39 new pairs.

The foreground quasars are distributed between $z\approx0.8$-$3.6$ and their halos are probed by 
their background counterparts over physical scales of 
$\mathrm{R_{\perp}}\approx 0.1$-$0.5\,\mathrm{Mpc}$. The coordinates, redshifts, on-sky and physical 
separations are given for the new quasar pairs in Table~\ref{tab:newpairs}. The star symbols in 
Figure~\ref{fig:colorplots} show the positions of the new projected quasars in $ugrizW1$ color-color 
space and are color coded according to their redshifts.

\startlongtable
\begin{deluxetable*}{ccccccc}
\tablewidth{0pc}
\tablecaption{Table of new quasar pairs discovered in combined ATLAS, SDSS and WISE imaging.
\label{tab:newpairs}}
\tablehead{
\colhead{$\mathrm{QSO_{fg}}$} & 
\colhead{$\mathrm{QSO_{bg}}$} &
\colhead{$z_\mathrm{fg}$} & 
\colhead{$z_\mathrm{bg}$} &
\colhead{$\theta\arcmin$} & 
\colhead{$R_{\perp}\ (\mathrm{pkpc})$} &
\colhead{Survey}
}
\startdata
J003308.63-083222.19&J003307.31-083241.55&3.038&3.043&0.459&216.433&SDSS \\
J012902.78+191824.46&J012901.92+191847.18&2.680&2.691&0.430&209.778&SDSS \\
J015415.22+032455.84&J015416.43+032457.86&2.660&3.219&0.304&148.640&SDSS \\
J022845.72-124643.92&J022848.07-124706.78&1.733&2.032&0.688&358.303&ATLAS\\
J023229.05-100123.48&J023231.25-100102.92&2.063&2.386&0.641&328.856&ATLAS\\
J031855.31-103040.30&J031853.87-102945.32&2.226&2.417&0.982&498.485&ATLAS\\
J032926.40-134732.22&J032926.04-134831.51&2.073&2.372&0.992&508.701&ATLAS\\
J033347.40-133928.44&J033345.40-133938.41&2.230&2.679&0.513&260.495&ATLAS\\
J034952.34-110620.59&J034955.77-110642.91&2.449&2.824&0.920&458.622&ATLAS\\
J090551.96+253003.35&J090551.25+253026.09&3.325&3.300&0.411&188.462&SDSS \\
J090828.30+080313.18&J090826.82+080320.34&2.390&3.168&0.385&193.025&SDSS \\
J091800.77+153621.46&J091800.70+153631.31&2.980&2.958&0.165& 78.282&SDSS \\
J093240.91+400905.65&J093243.02+400913.95&2.962&3.130&0.426&202.556&SDSS \\
J093836.78+100905.34&J093837.81+100922.00&2.504&2.818&0.376&186.529&SDSS \\
J095503.57+614242.66&J095503.14+614247.33&2.739&2.725&0.093& 45.173&SDSS \\
J095549.38+153838.11&J095549.80+153837.00&0.830&2.900&0.103& 48.214&SDSS \\
J095629.72+243441.34&J095627.88+243436.98&2.979&2.914&0.425&201.423&SDSS \\
J100205.70+462411.82&J100202.89+462407.25&3.138&2.760&0.490&228.948&SDSS \\
J100253.37+341924.03&J100254.22+341928.47&2.418&2.506&0.190& 95.194&SDSS \\
J100903.16-142104.27&J100859.11-142114.19&2.033&2.068&0.995&511.335&ATLAS\\
J101853.24-160727.80&J101853.10-160808.04&2.331&2.953&0.672&338.026&ATLAS\\
J102947.32+120817.11&J102945.77+120824.53&2.820&3.392&0.399&192.024&SDSS \\
J103109.37+375749.68&J103108.25+375801.19&2.752&2.589&0.292&141.842&SDSS \\
J103716.68+430915.57&J103716.86+430944.83&2.676&3.286&0.489&238.758&SDSS \\
J104314.33+143434.81&J104313.69+143435.73&2.980&3.361&0.156& 73.812&SDSS \\
J104339.12+010531.29&J104338.28+010507.77&3.240&3.001&0.445&205.465&SDSS \\
J105202.95-103803.70&J105203.23-103815.09&2.104&2.194&0.202&103.336&ATLAS\\
J105338.15-081623.66&J105336.09-081620.94&2.192&2.294&0.512&260.282&ATLAS\\
J105354.90-100941.44&J105354.48-100931.71&3.232&3.248&0.192& 88.924&ATLAS\\
J110402.08+132154.46&J110401.42+132134.70&2.869&2.576&0.366&175.702&SDSS \\
J110124.79-105645.12&J110126.03-105642.26&2.579&2.688&0.308&151.832&ATLAS\\
J111820.36+044120.22&J111820.46+044125.26&3.120&3.454&0.088& 40.981&SDSS \\
J112032.04-095203.21&J112032.65-095138.28&2.180&3.627&0.442&224.954&ATLAS\\
J112239.32+450618.54&J112236.72+450628.12&3.590&3.044&0.486&216.459&SDSS \\
J112355.97-125040.73&J112359.53-125056.76&2.965&3.428&0.908&431.314&ATLAS\\
J112516.06+284057.59&J112516.26+284122.74&2.845&2.834&0.421&202.590&SDSS \\
J112839.64-144842.36&J112843.30-144837.44&1.920&2.200&0.888&459.457&ATLAS\\
J112913.52+662039.13&J112915.28+662101.63&2.807&2.803&0.414&199.966&SDSS \\
J113820.28+203336.93&J113820.42+203333.18&2.687&2.679&0.071& 34.437&SDSS \\
J114443.59+102143.48&J114442.32+102125.21&1.503&2.833&0.436&227.342&SDSS \\
J115037.52+422421.01&J115035.53+422409.90&2.883&3.126&0.411&197.017&SDSS \\
J115222.15+271543.29&J115221.84+271540.80&3.102&3.083&0.080& 37.686&SDSS \\
J120032.34+491951.99&J120034.26+492015.22&2.629&3.254&0.498&244.193&SDSS \\
J121642.25+292537.97&J121641.77+292529.34&2.532&2.519&0.178& 87.996&SDSS \\
J122900.87+422243.23&J122859.36+422229.73&3.842&3.459&0.358&155.535&SDSS \\
J123055.78+184746.79&J123056.94+184736.83&3.169&3.089&0.321&149.312&SDSS \\
J132728.77+271311.96&J132729.83+271324.94&3.085&2.658&0.320&150.152&SDSS \\
J134221.26+215041.97&J134219.85+215051.20&3.062&2.506&0.362&170.098&SDSS \\
J135456.96+494143.74&J135456.76+494154.08&3.126&2.928&0.175& 81.962&SDSS \\
J141457.24+242039.67&J141457.12+242106.23&3.576&3.515&0.444&197.922&SDSS \\
J143622.50+424127.13&J143622.01+424132.22&3.000&3.050&0.124& 58.564&SDSS \\
J144225.30+625600.96&J144223.04+625625.99&3.271&3.271&0.490&225.685&SDSS \\
J162413.70+183330.72&J162412.59+183348.25&2.763&3.263&0.393&190.480&SDSS \\
J214858.11-074033.28&J214858.06-074034.98&2.660&2.660&0.031& 15.128&SDSS \\
\enddata
\tablecomments{From left to right columns give the names of the foreground and background 
  quasars, the foreground and background quasar redshifts, the on-sky angular separation 
  between the pair in arcminutes, the physical transverse distance between the line of sight
  of the background quasar and the foreground quasar in pkpc and the survey in which the pair
  was discovered.}
\end{deluxetable*}

\section{The Quasar pair database}\label{sec:database}

The database of quasar pairs comprises a catalog and a spectral library both housed within a single 
{\sc HDF5\ specDB} file. {\sc specDB} is a software 
package\footnote{\url{https://github.com/specdb/specdb}} for generating and interfacing with 
databases of astronomical spectra, written and maintained by JXP. The following sections focus on 
the database content, source and data characteristics and the database architecture. Full 
documentation of the {\sc specDB} package, including download instructions for the spectral 
database, are supplied by Read the Docs\footnote{\url{http://specdb.readthedocs.io/en/latest/}}. We 
further note that the {\sc igmspec} database, also under {\sc specDB}, provides approximately 
500,000 quasar spectra from public and private datasets \citep{2017A&C....19...27P}. In keeping with 
its goals, which include maintaining a highly complete database of quasar spectra in a consistent 
format, the quasar pair spectra will also be ingested into {\sc igmspec}, however some of the 
catalog content will only be available via the QPQ pair catalog presented here.

\subsection{The catalog}\label{sec:database:cat}

The quasar pair catalog comprises of a simple table containing a single record for each unique 
source. Distinct catalog records are uniquely identified via a primary key. The catalog also 
contains celestial coordinates, redshifts, references to the aforementioned, a redshift uncertainty 
column, which remains empty and UV and mid-IR photometry for all pairs within $5\arcmin$\ of 
separation. The celestial coordinates come largely from PanStarrs \citep{2016arXiv161205560C} 
because, except for a small fraction of objects that fall outside of the footprint, PanStarrs covers 
the entire catalog with sub-arcsecond accuracy.

The catalog redshifts are estimated via a wide range of distinct methodologies from a variety of 
sources including cross correlation \citep{2010MNRAS.405.2302H}, principle component analysis 
\citep{2012A&A...548A..66P}, spectral line fitting \citepalias{2013ApJ...776..136P} and visual 
inspection (this submission). The most significant difficulty in estimating quasar redshifts is in 
accounting for the natural variance in emission line properties of the quasar population both as a 
function of redshift and luminosity. This variation is large and ill-understood and some redshift 
estimators are better than others in accounting for it. Consolidating redshift uncertainties with 
differing systematics into a single table would be inconsistent and misleading. Instead redshift 
uncertainties have been estimated at $\delta\simeq\,1000\mathrm{km\,s^{-1}}$, which is 
conservative\footnote{Note also that \citet{2016ApJ...831....7S} give good general guidelines for 
the uncertainty associated with particular lines, which may be used to estimate the uncertainty at a 
given redshift, simply by considering which lines are redshifted into the optical window.}. The 
redshift uncertainty column is empty but remains in the catalog to facilitate future attempts to 
measure redshifts more consistently (see section ~\ref{sec:summary}). 

Columns of mid-IR and UV photometry are provided via measurements made at the positions of catalog 
objects in the unWISE coadds (section~\ref{sec:atlaswise}) and the {\it Galaxy Evolution Explorer} 
\citep[GALEX;][]{2007ApJS..173..682M} imaging products. Forced photometry from these two surveys is 
included for the reasons discussed in section~\ref{sec:atlaswise}. 

GALEX undertook wide field surveys in both imaging and low resolution grism spectroscopy from May 
2003 until February 2012. It delivered the first broadband imaging surveys in the far-UV and near-UV 
at central wavelengths of $1528$\AA\ and $2310$\AA\ respectively. 

Flux-calibrated, background-subtracted intensity maps as well as the sky background and threshold 
weight maps are served by the Barbara A.~Mikulski Archive for Space Telescopes (MAST). The GALEX 
photometry pipeline \citep{2007ApJS..173..682M} passes these products through {\sc SExtractor} 
\citep{1996A&AS..117..393B} to obtain calibrated catalogs. MAST also serves the {\sc SExtractor} 
configuration and parameter files necessary for computing photometery on each individual image. In 
principle then, one should be able to reconstruct any GALEX catalog in the official release. The 
slight complication with forced photometry is the need to bypass the source detection stage and 
simply place apertures down at predefined pixel positions. {\sc SExtractor} does not have this 
facility and therefore forced photometry was performed by constructing mock images with mock sources 
at the position of the object of interest. {\sc SExtractor} was then run in dual image mode, which 
allows the mock image to be used for the purposes of source detection and the real image to be used 
for the source extraction, thereby extracting the real source from the real image at the position of 
the mock source in the mock image. Beyond this modification the procedure follows that of the actual 
GALEX photometry pipeline. Photometry is provided for a 6\arcsec\ radius aperture, which is a 
reasonable compromise between minimizing background noise contributions and measuring photometry 
towards the field edges where the PSF becomes degraded.

\begin{figure}
   \centering
   \includegraphics[scale=0.5]{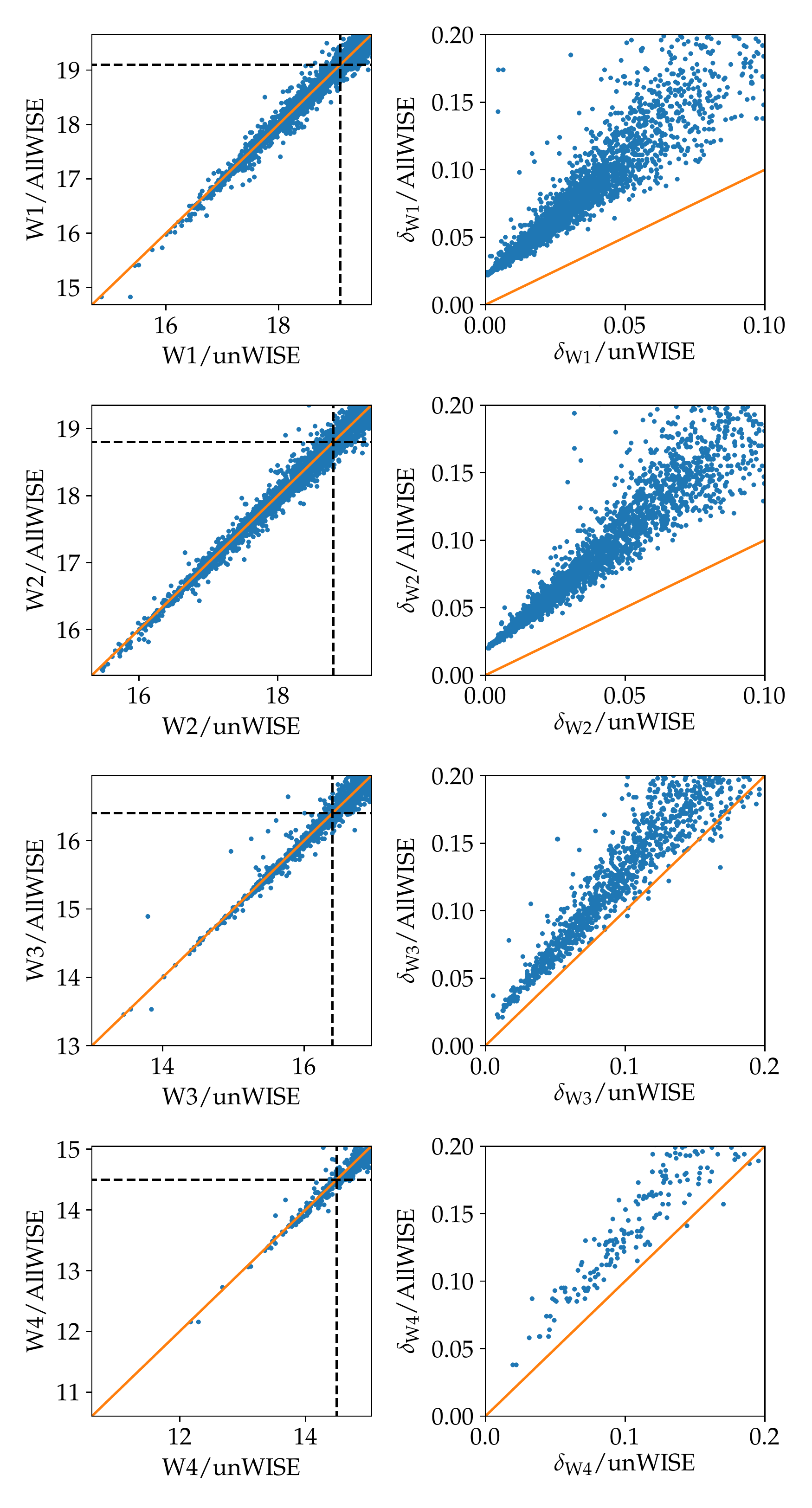}
   \caption{A comparison of the AllWISE photometry of detected quasar pairs with the 
   forced photometry of the unWISE images. Left had panels show the magnitude 
   measurements from the AllWISE release versus the forced measurements. The orange line 
   shows a one-to-one relationship and the dashed lines are placed at the average 
   $5\sigma$ AllWISE limiting magnitudes. The right hand panels compare the magnitude 
   errors. The gains in the unWISE over the AllWISE dataset are obviously apparent, 
   especially in the W1 and W2 bands where the addition of NEOWISE and NEOWISER data has 
   approximately doubled the signal-noise.}
   \label{fig:wisecompare}
\end{figure}

\begin{figure}
   \centering
   \includegraphics[scale=0.55]{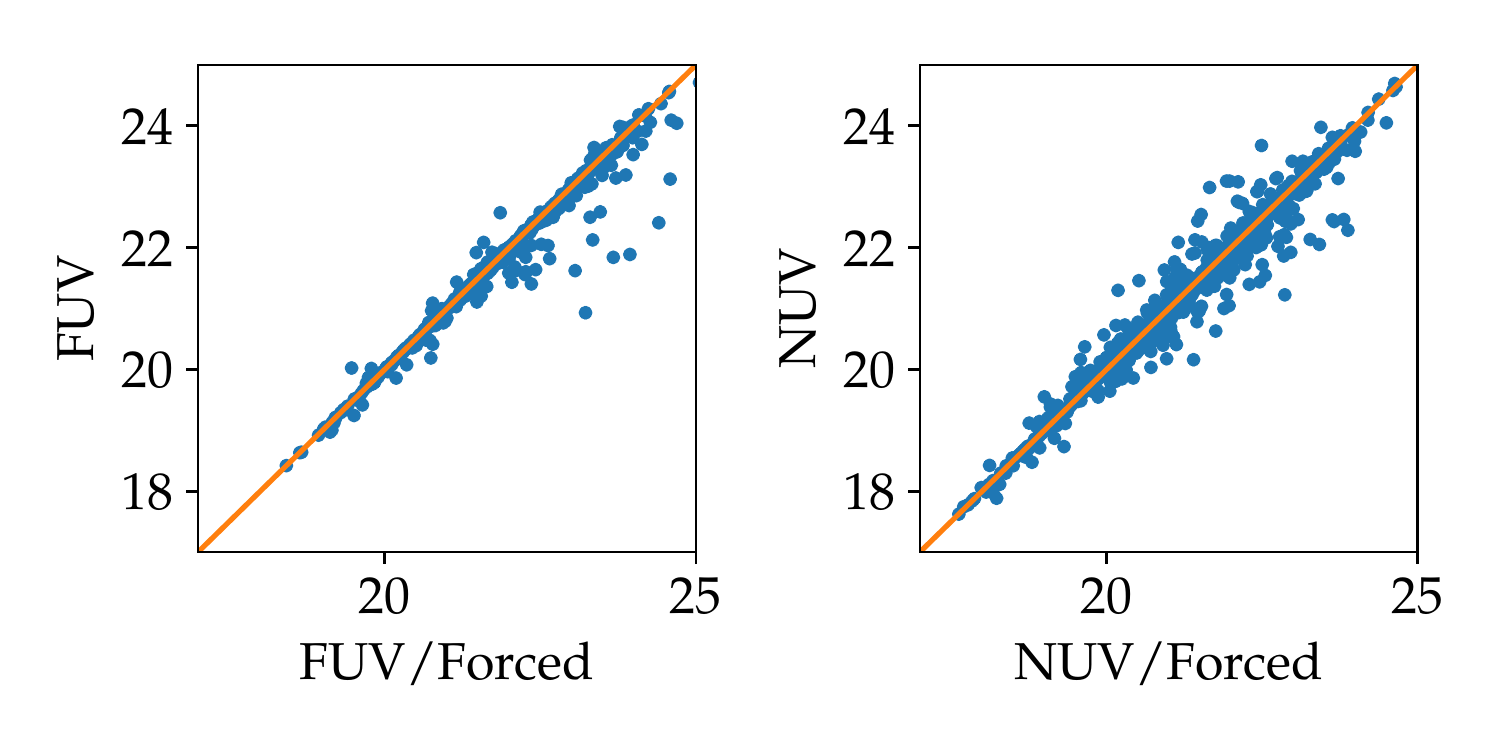}
   \caption{A comparison of officially released GALEX photometry for detected quasars 
   versus forced photometry. The left panel shows FUV photometry and the right panel
   shows NUV photometry. The orange line shows a one-to-one relationship and the two sets 
   of measurements are in good agreement, which serves to verify the forced photometry 
   method.}
   \label{fig:galexcompare}
\end{figure}

The results of the unWISE and GALEX forced photometry are verified in Figures~\ref{fig:wisecompare} 
and \ref{fig:galexcompare} respectively, where members of the quasar pair catalog which are detected 
in the officially released AllWISE or GALEX catalogs are plotted against the forced photometry. In 
Figure~\ref{fig:wisecompare} the left-hand panels show the comparison of AllWISE and unWISE 
photometry and the orange lines show a one-to-one relationship. The plot axes extend to the average 
$3\sigma$ depth of the official release in each band and the dashed lines show the average $5\sigma$ 
depth in each band. The right-hand panels compare magnitude errors. The benefits of using the unWISE 
products over the official AllWISE release are clear. Significant gains in signal-noise are achieved 
in all bands and particularly in W1 and W2 where the addition of the NEOWISE observations have 
almost doubled the signal-noise over the AllWISE release.

Figure~\ref{fig:galexcompare} shows a similar comparison between the officially released and forced 
GALEX photometry. The forced photometry in both cases, unWISE and GALEX, is in good agreement with 
their respective official releases. Forced photometry of quasar pairs in optical or near-IR surveys 
is omitted from the catalog since straight-forward catalog matching is both efficient and accurate 
at the fine spatial resolutions offered by modern surveys in these wavelength regimes.

\begin{figure*}
   \centering
   \includegraphics[scale=0.75]{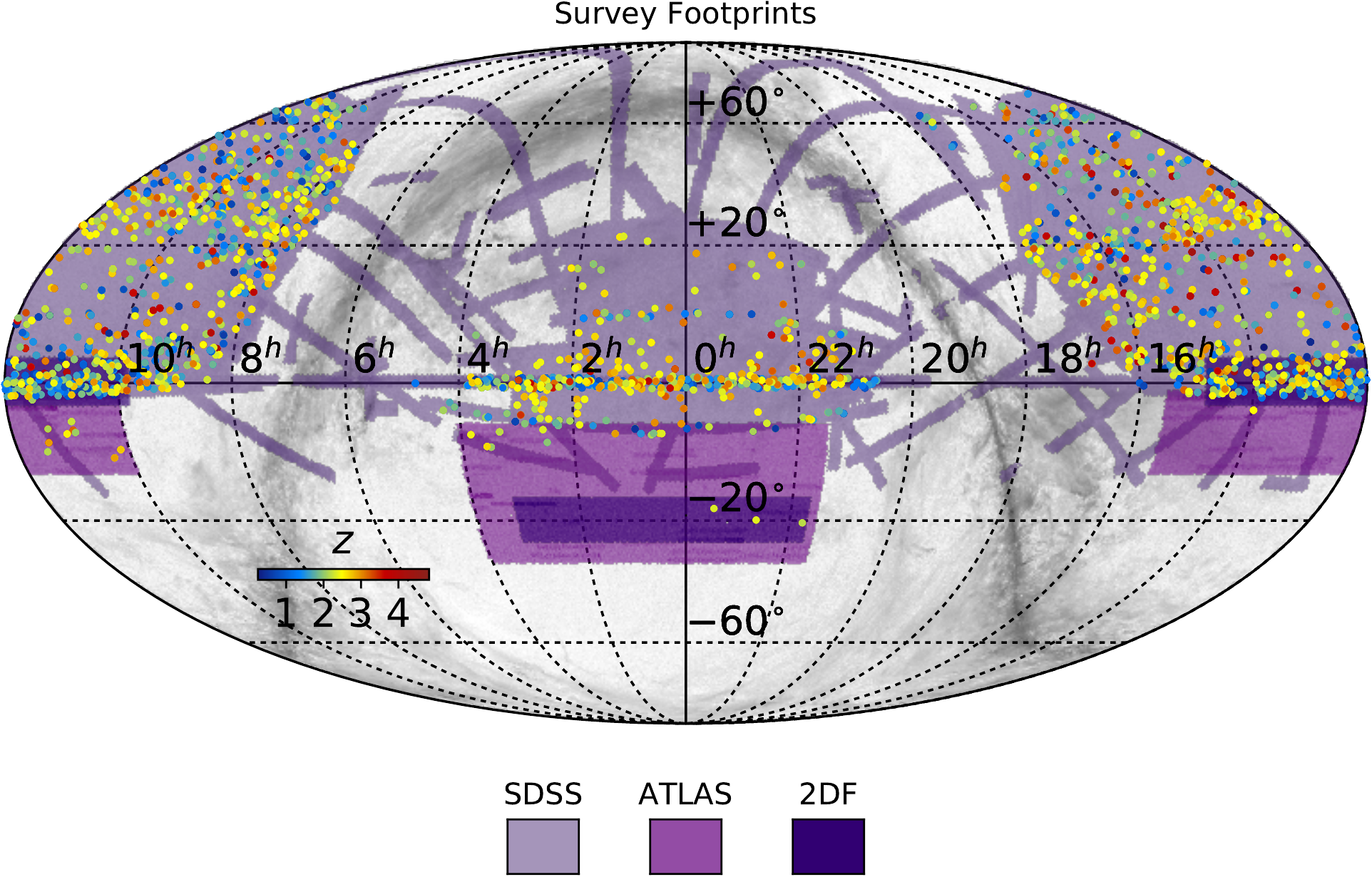}
   \caption{The sky coverage of the QPQ catalog in equatorial coordinates. Points show the 
   locations of all foreground quasar pair members and are color coded according to 
   redshift. The shaded regions show the various survey footprints which bound the QPQ 
   search area and are color coded as given in the legend. The SDSS-LS and BOSS imaging 
   footprints are plotted in the same color labeled ``SDSS". The gray scale background 
   shows the Milky Way polarized dust emission as seen by {\it Planck}.}
   \label{fig:skyplot}
\end{figure*}

A full description of the catalog is given in Table~\ref{tab:catalog}. Figure \ref{fig:skyplot} 
shows the sky coverage of the catalog with points plotted at the locations of all foreground quasars 
and color coded according to redshift. The filled regions correspond to the imaging footprints of 
the surveys that bound the QPQ quasar searches, namely SDSS-LS, BOSS, ATLAS and 2QZ. The background 
map shows the Milky Way polarized dust emission from the {\it Planck} commander component separation 
\citep{2015A&A...576A.104P}. 

\begin{figure}
   \centering
   \includegraphics[scale=0.7]{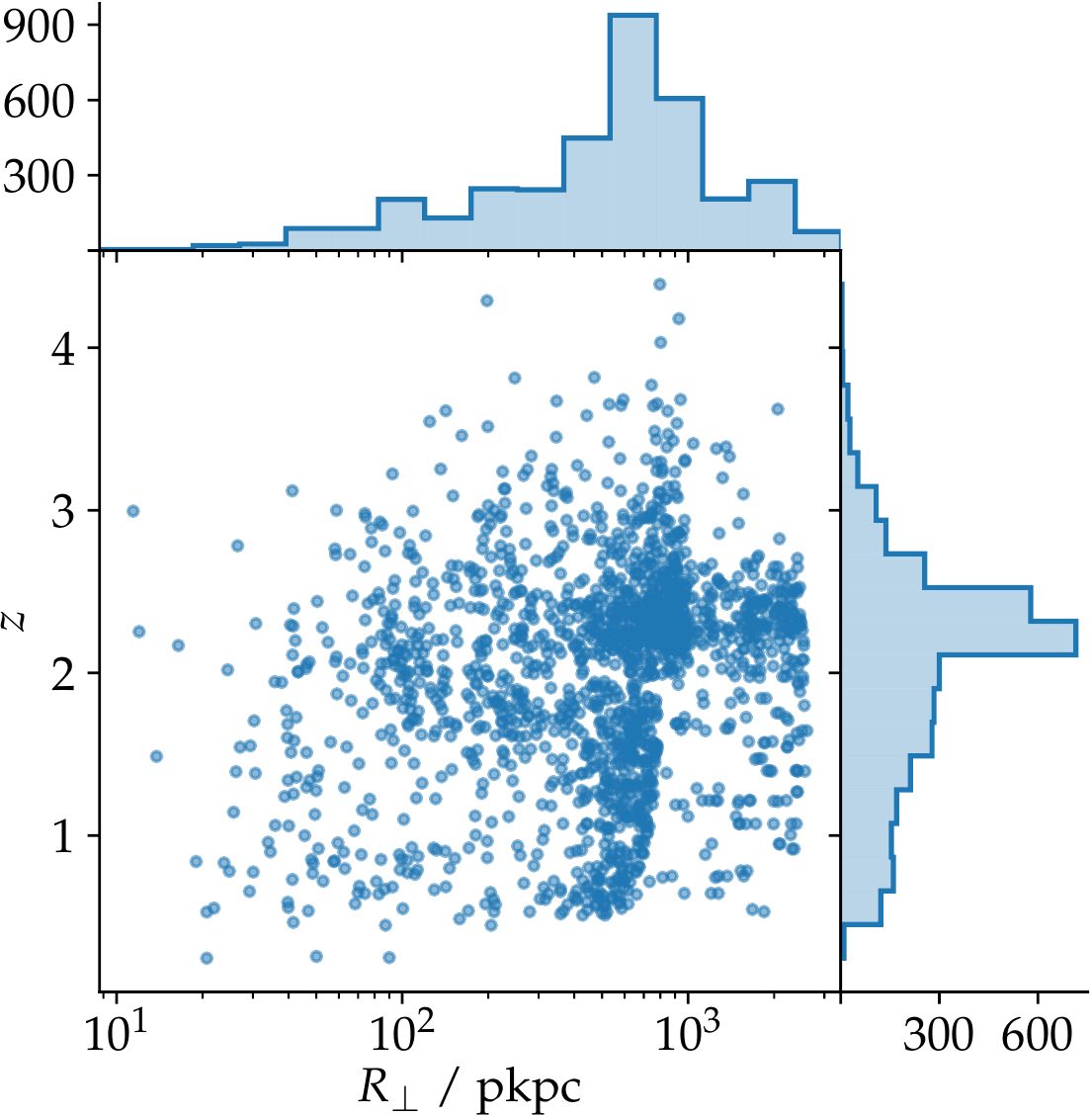}
   \caption{The joint distribution of redshift and physical transverse separation       
   $R_{\perp}$ is shown in the main panel. Each is also plotted independently in the 
   histograms on the peripheral panels. The peak in the redshift distribution corresponds
   to the peak in cosmic AGN activity. The distribution of physical transverse separation
   shows that the catalog probes the halos of quasars at impact parameters of tens of 
   physical kpc to a few Mpc. Physical binaries are omitted in each of these plots
by cutting the sample to pairs with velocity differences of $>3000\,\mathrm{km\,s^{-1}}$.}
   \label{fig:physsep}
\end{figure}

The distribution of foreground quasar redshifts is shown in the right hand panel of the joint-plot 
in Figure \ref{fig:physsep}. The distribution peaks at $z\sim 2.5$, corresponding to the peak in 
cosmic quasar activity. From the redshifts of foreground and background pair members and the known 
angular separations between them, one can compute the proper transverse separation of a foreground 
quasar to the sightline of the background quasar. The distribution of proper transverse separations 
is shown in the upper panel of Figure~\ref{fig:physsep} and the joint distribution with redshift is 
shown in the main panel of the same Figure. Physical binaries are omitted in each of these plots by 
cutting the sample to pairs with velocity differences of $>3000\,\mathrm{km\,s^{-1}}$. This plot 
serves to illustrate the range of environments probed by the background quasar sightlines. Distances 
of tens of $\mathrm{pkpc}$, corresponding to the outer regions of galactic discs to hundreds of 
$\mathrm{pkpc}$, probing the CGM, to a few $\mathrm{pMpc}$ corresponding to scales in the cosmic web 
are all probed by the background sources.

\subsection{The spectroscopic library}

The spectroscopic library houses the spectra of quasar pairs in the catalog listing. There may be 
multiple spectra associated with any distinct catalog source. The spectroscopic library is a 
heterogeneous data set which includes low to moderate to high resolution spectra with wavelength 
coverage from the optical to the near-IR. The low and many of the moderate resolution spectra 
generally result from QPQ campaigns focused on fast and efficient spectroscopic identification of 
photometric targets. A significant fraction of the optical, moderate-resolution spectra come from 
SDSS-LS or BOSS. The high resolution spectra were specifically targeted towards and have been used 
in detailed studies of the CGM \citepalias{2009ApJ...690.1558P,2017arXiv170503476L}. 

Over the years, related projects with broadened science goals have extended the QPQ catalog further. 
The various science cases have included measuring the small-scale clustering of quasars 
\citep{2006AJ....131....1H,2007ApJ...658...99M,2008ApJ...678..635M,2010ApJ...719.1672H,
2010ApJ...719.1693S,2017MNRAS.468...77E}, exploring correlations in the IGM along close separation 
sightlines \citep{2007MNRAS.378..801E,2010ApJ...721..174M}, analyzing small-scale transverse 
Ly$\alpha$ forest correlations \citep{2013ApJ...775...81R}, characterizing the transverse proximity 
effect \citep{2017ApJ...847...81S}, probing the halos of damped Ly$\alpha$ systems 
\citep[DLAs;][]{2015ApJ...808...38R}and correcting CIV-based virial BH masses
\citep{2017MNRAS.465.2120C}

The latter submission describes the near-infrared spectra of approximately 120 quasar pairs observed 
as part of QPQ follow-up programs. \citeauthor{2017MNRAS.465.2120C} also provide an additional ~500 
near-IR quasar spectra of non-pairs compiled both from the literature and from their own 
observations. At the time of writing all the \citeauthor{2017MNRAS.465.2120C} near-IR spectra are 
restricted to proprietary use but are expected to become publicly available in the very near future 
\citetext{P.~C.\ Hewett priv.~comm.}. As soon as this occurs, the near-IR pair spectra will be 
ingested by the QPQ database\footnote{Non-pair spectra will be ingested by the {\it igmspec} 
database, also under the SpecDB software package.}.

\begin{figure}
   \centering
   \includegraphics[scale=0.65]{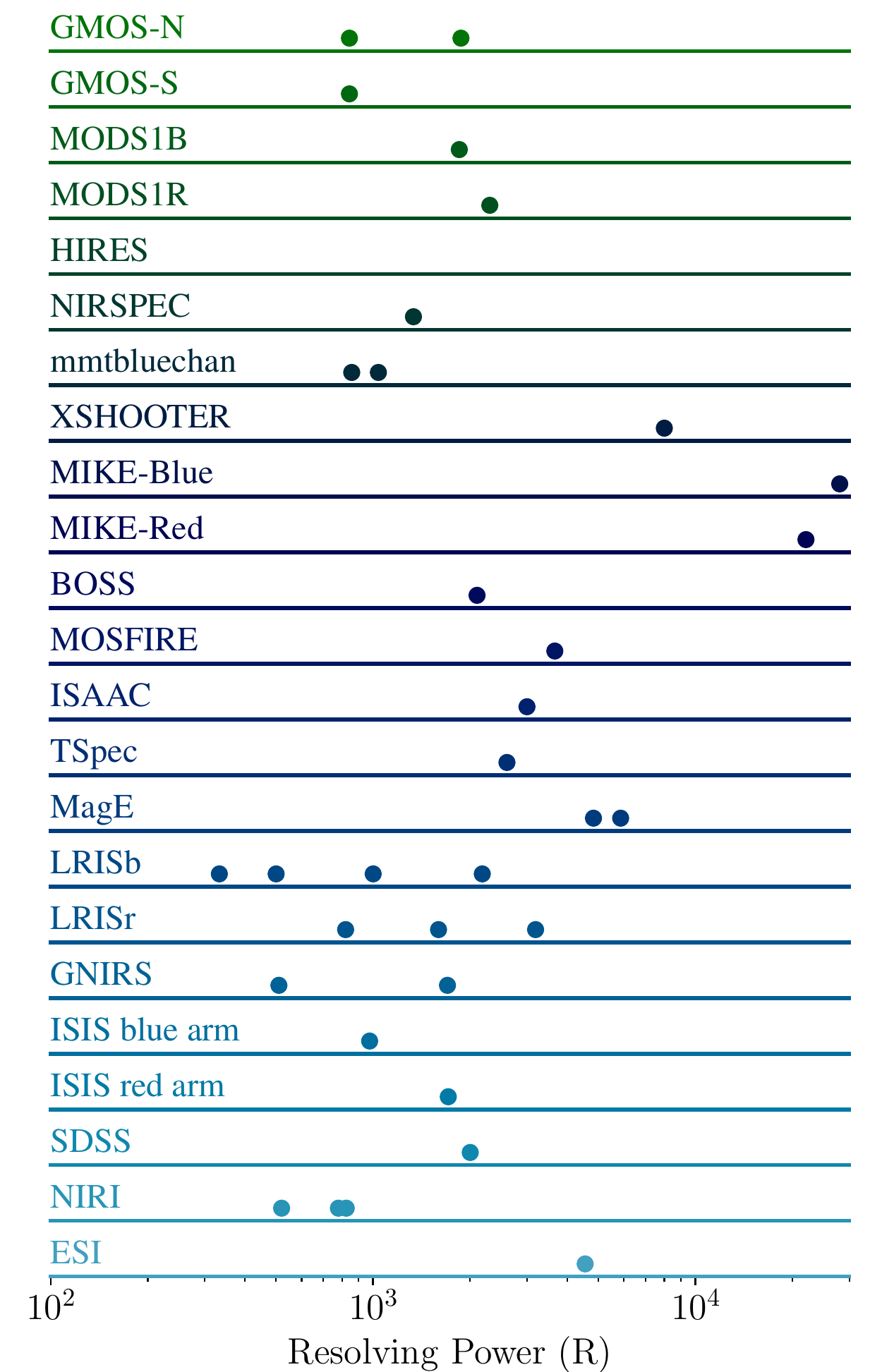}
   \caption{The range in spectral resolving power grouped by instrument. Resolving power (R) is 
   given along the $x$-axis for each instrument listed on the $y$-axis. Note that no 
   numerical scale is associated with the $y$-axis and that this figure does not give 
   information pertaining to the number of spectra associated with a given instrument
   (see Table~\ref{tab:wve}). Rather unique R-values for all spectra associated with a 
   given instrument are represented by single points along the $x$-axis.}
   \label{fig:resolution}
\end{figure}

The range in resolving power covered by the spectral library is shown for each instrument in 
Figure~\ref{fig:resolution}. The wavelength coverage of the spectra are characterized in 
Table~\ref{tab:wve}, with respect to the optical and near-IR broad-band filters of the SDSS and 
UKIDSS imaging surveys. Each row corresponds to a particular telescope and instrument in the 
catalog. Column ``Total'' refers to the total number of spectra. The columns $ugriz$YJHK refer to 
the SDSS or UKIDSS passbands of the same name and indicate the number of spectra with coverage in 
those passpands. A spectrum is arbitrarily considered to have coverage in a given passband when its  wavelength array falls entirely or partially within the cut-on and cut-off wavelengths at 50 percent 
transmission. To avoid counting the less useful low signal-noise regions of any spectrum (usually 
found towards detector edges), positive wavelength coverage also requires that the average 
signal-noise in the 50 pixels either side of the central covering pixel is at least 3. 

\begin{deluxetable*}{llcccccccccc}
\tablewidth{0pc}
\tablecaption{Wavelength coverage
\label{tab:wve}}
\tablehead{
\colhead{Telescope} & 
\colhead{Instrument} &
\colhead{Total} &
\colhead{$u$} &
\colhead{$g$} &
\colhead{$r$} &
\colhead{$i$} &
\colhead{$z$} &
\colhead{$Y$} &
\colhead{$J$} &
\colhead{$H$} &
\colhead{$K$}
}
\startdata
Gemini-North&GMOS-N&71&28&70&37&0&0&0&0&0&0\\
Gemini-North&NIRI&33&0&0&0&0&0&0&0&30&3\\
Gemini-South&GMOS-S&36&3&34&36&6&0&0&0&0&0\\
Gemini-South&GNIRS&29&0&0&0&2&24&28&29&28&29\\
MGIO-LBT&MODS1B&10&7&10&0&0&0&0&0&0&0\\
MGIO-LBT&MODS1R&10&0&0&10&10&10&0&0&0&0\\
Keck-I&LRISb&358&258&216&129&0&0&0&0&0&0\\
Keck-I&LRISr&140&0&0&136&136&134&0&0&0&0\\
Keck-I&HIRES&1&1&1&1&0&0&0&0&0&0\\
Keck-I&MOSFIRE&2&0&0&0&0&0&0&1&2&0\\
Keck-II&NIRSPEC&5&0&0&0&0&0&0&0&2&2\\
Keck-II&ESI&115&0&86&109&112&108&89&0&0&0\\
MMTO&mmtbluechan&91&68&87&22&0&0&0&0&0&0\\
ESO-VLT-U2&XSHOOTER&36&15&32&33&32&29&23&33&34&8\\
ESO-VLT-U3&ISAAC&17&0&0&0&0&0&0&0&17&0\\
Clay (Mag.II)&MIKE-Blue&6&2&5&2&2&2&0&0&0&0\\
Clay (Mag.II)&MIKE-Red&4&0&4&4&4&2&0&0&0&0\\
Clay\_Mag\_2&MagE&76&32&73&75&75&74&39&0&0&0\\
SDSS 2.5-M&SDSS&88&46&72&83&83&57&0&0&0&0\\
SDSS 2.5-M&BOSS&2304&526&1896&1539&1526&1305&221&0&0&0\\
200&TSpec&68&0&0&0&0&0&64&67&67&52\\
WHT&ISIS blue arm&41&0&23&0&0&0&0&0&0&0\\
WHT&ISIS red arm&41&0&0&35&29&0&0&0&0&0\\
\enddata
\tablecomments{
Wavelength coverage of individual spectra in the spectral 
library. Wavelength coverage is described with respect to
the $ugrizYJHK$ bandpasses of the SDSS and UKIDSS imaging
surveys. A spectrum is arbitrarily defined to have coverage
in a particular bandpass if its wavelength array falls within
the interval of the cut-on cut-off values at 50 per cent of
the peak bandpass transmission. In order to avoid counting
low signal-noise regions of the spectrum, the average
signal-noise in the 50 pixels either side of the central
covering pixel must have signal-noise of at least 3. The column
``Total'' gives the total number of spectra grouped according to a
particular Telescope and instrument combination. The columns
$ugrizYJHK$ give the number of spectra deemed to have coverage
in that bandpass.}
\end{deluxetable*}

\subsection{Database Architecture}

The database comprises a catalog listing and a spectral library. The catalog listing is a simple 
table containing one record for each unique source in the database. Each field in the catalog is 
described in Table~\ref{tab:catalog}. The spectral library may contain one or more spectra for each 
individual catalog source. A catalog record is linked to its corresponding spectra via a primary key 
field unique at the catalog level and having a one to many relationship with all spectra associated 
to that source.

{\catcode`\&=11
\gdef\2012AandA...548A..66P{\cite{2012A&A...548A..66P}}}

\begin{deluxetable*}{lll}
\tablewidth{0pc}
\tablecaption{Catalog schema.
\label{tab:catalog}}
\tablehead{
\colhead{Key} & 
\colhead{Type} &
\colhead{Depscription}
}
\startdata
{\tt QPQ\_ID}&{\tt INT}&Primary key unique identifier\\
{\tt flag\_group}&{\tt INT}&{Bitwise flag indicating the groups that the source has spectra in}\\
{\tt zem}&{\tt FLOAT}&{Emission redshift of source}\\
{\tt sig\_zem}&{\tt FLOAT}&{Estimated error in the redshift}\tablenotemark{a}\\
{\tt flag\_zem}&{\tt STR}&{Key indicating source of the redshift}\tablenotemark{b}\\
{\tt RA}&{\tt FLOAT}&{Celestial Right Ascension in decimal degrees}\\
{\tt DEC}&{\tt FLOAT}&{Celestial declination in decimal degrees}\\
{\tt flag\_coo}&{\tt STR}&{Key indicating source of the coordinates}\\
{\tt STYPE}&{\tt STR}&{Spectral type (e.g. QSO)}\\
{\tt W1\_FLUX}&{\tt FLOAT}&WISE W1 AB flux in nanomaggies\\
{\tt W2\_FLUX}&{\tt FLOAT}&WISE W2 AB flux in nanomaggies\\
{\tt W3\_FLUX}&{\tt FLOAT}&WISE W3 AB flux in nanomaggies\\
{\tt W4\_FLUX}&{\tt FLOAT}&WISE W4 AB flux in nanomaggies\\
{\tt W1\_IVAR}&{\tt FLOAT}&WISE W1 AB invserse variance in nanomagies$^{-2}$\\
{\tt W2\_IVAR}&{\tt FLOAT}&WISE W2 AB invserse variance in nanomaggies$^{-2}$\\
{\tt W3\_IVAR}&{\tt FLOAT}&WISE W3 AB invserse variance in nanomaggies$^{-2}$\\
{\tt W4\_IVAR}&{\tt FLOAT}&WISE W4 AB invserse variance in nanomaggies$^{-2}$\\
{\tt FUV\_FLUX}&{\tt FLOAT}&GALEX FUV AB flux in nanomaggies\\
{\tt NUV\_FLUX}&{\tt FLOAT}&GALEX NUV AB flux in nanomaggies\\
{\tt FUV\_IVAR}&{\tt FLOAT}&GALEX FUV AB invserse variance in nanomaggies$^{-2}$\\
{\tt NUV\_IVAR}&{\tt FLOAT}&GALEX NUV AB invserse variance in nanomaggies$^{-2}$\\
\enddata
\tablenotetext{a}{Redshift uncertainties are currently set to zero (see 
section~\ref{sec:database:cat}) }
\tablenotetext{b}{Possibilities are HW2010:\ \citet{2010MNRAS.405.2302H}, BOSS\_PCA:\ 
\2012AandA...548A..66P, QPQ:\ \citetalias{2006ApJ...651...61H}-
\citetalias{2017arXiv170503476L} and this submission.}
\end{deluxetable*}

Within the spectral library, spectra are arranged within a set of distinct groups. Each group 
contains the spectra and a meta-data table, which maintains a list of the common properties 
pertaining to each spectrum, including the primary key, wavelength coverage, resolving power, 
telescope and instrument etc. Entries in the meta-data table are ordered identically and aligned 
row-by-row with their corresponding spectra. A complete description of the meta-data fields is given 
in Table~\ref{tab:meta}. 

\begin{deluxetable}{lll}
\tablewidth{0pc}
\tablecaption{Meta-data schema.
\label{tab:meta}}
\tablehead{
\colhead{Key} & 
\colhead{Type} &
\colhead{Depscription}
}
\startdata
{\tt QPQ\_ID}&{\tt INT}&{Primary key unique at the catalog level}\\
{\tt GROUP\_ID}&{\tt INT}&{Primary key unique at the group level}\\
{\tt IGM\_ID}&{\tt INT}&{Primary key reference in to {\tt igmspec}}\\
{\tt zem\_GROUP}&{\tt FLOAT}&{Emission redshift}\\
{\tt sig\_zem}&{\tt FLOAT}&{Estimated error in the redshift}\\
{\tt flag\_zem}&{\tt STR}&{Key indicating source of the redshift}\\
{\tt RA\_GROUP}&{\tt FLOAT}&{Right Ascension in decimal degrees}\\
{\tt DEC\_GROUP}&{\tt FLOAT}&{Declination in decimal degrees}\\
{\tt EPOCH}&{\tt FLOAT}&{Year of epoch}\\
{\tt R}&{\tt FLOAT}&{Spectral resolution ($\delta \lambda/\lambda$): FWHM}\\
{\tt WV\_MIN}&{\tt FLOAT}&{Minimum wavelength value in \AA}\\
{\tt WV\_MAX}&{\tt FLOAT}&{Maximum wavelength value in \AA}\\
{\tt NPIX}&{\tt INT}&{Number of pixels in the spectrum}\\
{\tt SPEC\_FILE}&{\tt STR}&{Individual filename of the spectrum}\\
{\tt STYPE}&{\tt STR}&{Spectral type (e.g. QSO)}\\
{\tt INSTR}&{\tt STR}&{Instrument}\\
{\tt DISPERSER}&{\tt STR}&{Dispersing element}\\
{\tt TELESCOPE}&{\tt STR}&{Name of the telescope}\\
{\tt GROUP}&{\tt STR}&{Name of group}\\
{\tt DATE-OBS}&{\tt STR}&{Observation date}\\
\enddata
\end{deluxetable}

The groups themselves are assigned and named according to the spectrograph used to measure the 
spectrum. A list of the groups can be found in Table~\ref{tab:groups}. Note that in most cases the 
name of the group is identical to the corresponding value in the instrument meta-data field. 
However, there are a few instances where this is not the case. In particular where a set of 
observations on a single spectrograph result in a blue channel and a red channel spectrum and these 
spectra have not been merged, then two instruments exist inside a single group, one for the red 
channel and one for the blue channel. 

\begin{deluxetable}{lll}
\tablewidth{0pc}
\tablecaption{Groups within the spectral library.
\label{tab:groups}}
\tablehead{
\multicolumn{3}{c}{Group Names}
}
\startdata
{\tt GMOS}&{\tt ISAAC}&{\tt MODS}\\
{\tt TRIPLESPEC}&{\tt HIRES}&{\tt MAGE}\\
{\tt NIRSPEC}&{\tt LRIS}&{\tt MMT}\\
{\tt GNIRS}&{\tt XSHOOTER}&{\tt SDSS}\\
{\tt MIKE}&{\tt NIRI}&{\tt BOSS}\\
{\tt ESI}&{\tt MOSFIRE}&{\tt ISIS}\\
\enddata
\end{deluxetable}

Given this simple architecture it is straightforward to pull spectra out of the library given a 
catalog search. For example one may query the catalog on any number of its fields to obtain a 
subsidiary table built to the query constraints. The primary key field of the subsidiary table can 
then be used to query the spectral library and retrieve the desired spectra. Of course, the catalog, 
spectral library and meta-data can also be queried independently of one another. 

\section{Summary \& Future Work}\label{sec:summary}

With the addition here of 54 newly discovered quasar pairs from VST ATLAS, SDSS and WISE, the QPQ 
database contains catalog listings for over 5500 distinct objects and a spectral database containing 
over 3500 optical and near-infrared spectra of projected quasar pairs, quasars closely separated in 
redshift and gravitational lens candidates. The database is the fruit of over a decade of work, nine 
previously published articles and many other related projects and studies. The projected pairs 
provide a means to probe the $z>2$ CGM of quasar host galaxies at impact parameters of tens of 
$\mathrm{pkpc}$ through to several $\mathrm{pMpc}$ or equivalently from scales comparable in extent 
to galactic discs, to bound gas in the CGM and to nearby regions of intergalactic space. In 
publishing this catalog the hope is to provide a laboratory for future discoveries in the CGM of 
massive galaxies hosting quasars. This database serves as a living resource which will continue to 
grow, reflecting advances in both scientific understanding and instrumentation.

New multi-fiber spectrographs such as DESI 
\citep[Dark Energy Spectroscopic Instrument;][]{2016arXiv161100036D} and Suburu PSF 
\citep[Prime Focus Spectrograph;][]{2016SPIE.9908E..1MT} will supply the data sets for future 
catalog expansion. DESI alone will target and obtain redshifts for over $\sim 700,000$ quasars at 
$z\gtrsim2$, providing gains of over 3 times in comparison to the combined SDSS, BOSS and the 
ongoing eBOSS quasar redshift surveys. The QPQ project lays the foundation for future pair searches 
in these data sets as well as the techniques to be able to study them in unprecedented and exquisite 
statistical detail.

The promise of future large spectroscopic surveys demands increasing numbers of parallel, detailed 
case studies. To that end, the pursuit of a much larger sample of high-resolution, high signal-noise 
spectra is of prime importance. The complexities of the CGM are manifest in its rich multiphase, 
multiscale structures, which display distinct kinematics and metallicities. The detailed dissection 
of all facets of the CGM requires the capability to resolve its smallest coherent structures. With 
current ground-based 10\,m telescopes few QPQ pairs are currently within range of echelle 
spectrographs which can provide resolutions of $\mathrm{FWHM}\sim 10\,\mathrm{km\,s^{-1}}$. The 
arrival of 30\,m class telescopes in the near future will place many of the QPQ pairs in the realm 
of these instruments and thus provide the required samples of high resolution spectra.

High resolution spectra are not useful in assessing the kinematics of distinct clouds or flows if 
line centroids cannot be measured with appreciable accuracy. This requires access to the HI Balmer 
series or narrow forbidden lines such as [OII] and [OIII], which at $z\sim 2$ are redshifted into 
the near-IR. In order to refine current kinematic constraints and provide the foundation for future 
high resolution observations, near-IR spectroscopy of quasar pairs is required. A campaign of near-
IR spectroscopy has been undertaken as part of the QPQ project, the spectra themselves are released 
in the database presented here, precise redshift measurement from these spectra will be presented in 
a forthcoming paper (Hennawi et al 2018 in prep.) and near-IR spectroscopic follow-up of quasar 
pairs continues.

In comparison to the success in cataloging projected quasar pairs, the pursuit of galaxies at small 
impact parameters from bright quasar sightlines has been less fruitful. By the same tenet, these 
projections are required to probe the CGM of ``normal'' galaxies. Despite over a decade of searches 
on 10\,m telescopes, only $\sim 10$ such sightlines currently exist with projected separations 
$\sim 200\,\mathrm{pkpc}$. On the other hand, there is strong evidence to suggest that Lyman Limit 
Systems (LLS), which are easily detected in quasar absorption spectra, originate in galactic halos 
\citep[e.g.][]{2013ApJ...775...78F,2016MNRAS.462.1978F}. When LLSs are captured in the absorption 
spectra of two more closely separated quasars, one can use the LLS autocorrelation function along 
these multiple close sightlines to glean the extent, covering factor and spatial profile of cool gas 
in the CGM.

It is similarly possible to study the interaction between different gas phases in the CGM by 
concentrating of intervening metal transitions. This experiment will elucidate the interplay between 
inflowing gas, expected to be metal poor, and outflows, which will be enriched. Such experiments are 
well underway, QPQ affiliated projects are using the $z\sim 3$ projected quasars presented in this 
submission and a sample of $z\sim 2$ quasar pairs, which have recently been observed by HST with the 
WFC3/UVIS grism to study the correlation of LLSs across the epoch of peak galaxy formation.

In contrast to well established techniques in absorption spectroscopy, the capacity to detect 
diffuse extragalactic gas in emission has been lacking until relatively recently. Cutting-edge 
techniques and advanced instruments have delivered some promising results and are set to alter this 
situation dramatically. Using custom built narrow-band filters to image quasar fields, several 
authors have reported the presence of Enormous Ly$\alpha$ nebulae (ELAN), illuminated by elevated UV 
radiation fields \citep{2014Natur.506...63C,2015Sci...348..779H,2014ApJ...786..106M}. Such are their 
projected angular sizes, ELAN are expected to extend well beyond the virial radii of quasar host 
galaxies, indicating that the emitting gas belongs to the surrounding IGM. This provides a new 
opportunity to study the characteristics of the gas feeding galaxies in emission and offers an 
independent and complimentary probe to absorption studies. Integral field unit spectrographs such as 
MUSE \citep[Multi Unit Spectroscopic Explorer][]{2010SPIE.7735E..08B}, CWI 
\citep[Cosmic Web Imager;][]{2010SPIE.7735E..0PM}, KCWI 
\citep[Keck Cosmic Web Imager;][]{2016SPIE.9908E..54R} and comparable instruments on 30\,m class 
telescopes will begin to lead this field. 

\citet{2015Sci...348..779H} demonstrated that physical quasar pairs may be signposts of ELAN. They 
report on the discovery of an ELAN in the presence of a physically associated quasar quartet. The 
chances of stumbling upon such a system serendipitously are $\sim 10^{-7}$, which strongly suggests 
a physical connection between ELAN and multiple quasars in overdense systems such as proto-clusters. 
QPQ physical quasar pair fields provide ideal locations for current and future searches for ELAN.

\acknowledgments
JRF and ADM acknowledge support from NSF grants 1515404 and 1616168, NASA grant 
NNX16AN48G and by the Director, Office of Science, Office of High Energy Physics of the 
U.S. Department of Energy under Contract No. DE- AC02-05CH1123. 

JXP and MWL acknowledge support from the National Science Foundation (NSF) grants AST-1010004 
and AST-1412981. MF acknowledges support by the Science and Technology Facilities Council 
grant number ST/P000541/1

The authors gratefully acknowledge the support which enabled observations on the WHT.
The WHT is operated on the island of La Palma by the Isaac Newton Group of Telescopes in the 
Spanish Observatorio del Roque de los Muchachos of the Instituto de Astrof\'isica de Canarias.
The WHT/ISIS spectroscopy was obtained as part of programs 2015/P1 and 2016/P6. We also 
acknowledge the support which enabled past observations at the Keck, Gemini, Large Binocular 
Telescope, Very Large Telescope, Las Campanas, MMT, Magellan, William Herschel Telescope, 
Palomar and Apache Point Observatories

This work was based on data products from observations made with ESO Telescopes at 
the La Silla Paranal Observatory under program ID 177.A-3011(A,B,C,D,E.F,G,H,I,J),
We also acknowledge support from Science and Technology Facilities Council
Consolidated Grant ST/P 000541/1.

Funding for the SDSS, SDSS-II and SDSS-III has been provided by the Alfred P.\ Sloan 
Foundation, the Participating Institutions, the National Science Foundation, the U.S. 
Department of Energy. SDSS and SDSS-II were additionally funded by the National Aeronautics 
and Space Administration, the Japanese Monbukagakusho, the Max Planck Society, and the Higher 
Education Funding Council for England. The SDSS Web Site is http://www.sdss.org/ the SDSS-III
website is http://www.sdss3.org/.

The SDSS is managed by the Astrophysical Research Consortium for the Participating 
Institutions. The Participating Institutions are the American Museum of Natural History, 
Astrophysical Institute Potsdam, University of Basel, University of Cambridge, Case Western 
Reserve University, University of Chicago, Drexel University, Fermilab, the Institute for 
Advanced Study, the Japan Participation Group, Johns Hopkins University, the Joint Institute 
for Nuclear Astrophysics, the Kavli Institute for Particle Astrophysics and Cosmology, the 
Korean Scientist Group, the Chinese Academy of Sciences (LAMOST), Los Alamos National 
Laboratory, the Max-Planck-Institute for Astronomy (MPIA), the Max-Planck-Institute for 
Astrophysics (MPA), New Mexico State University, Ohio State University, University of 
Pittsburgh, University of Portsmouth, Princeton University, the United States Naval 
Observatory, and the University of Washington.

SDSS-III is managed by the Astrophysical Research Consortium for the Participating 
Institutions of the SDSS-III Collaboration including the University of Arizona, the Brazilian 
Participation Group, Brookhaven National Laboratory, Carnegie Mellon University, University 
of Florida, the French Participation Group, the German Participation Group, Harvard 
University, the Instituto de Astrofisica de Canarias, the Michigan State/Notre Dame/JINA 
Participation Group, Johns Hopkins University, Lawrence Berkeley National Laboratory, Max 
Planck Institute for Astrophysics, Max Planck Institute for Extraterrestrial Physics, New 
Mexico State University, New York University, Ohio State University, Pennsylvania State 
University, University of Portsmouth, Princeton University, the Spanish Participation Group, 
University of Tokyo, University of Utah, Vanderbilt University, University of Virginia, 
University of Washington, and Yale University.

This publication makes use of data products from the Wide-field Infrared Survey Explorer, 
which is a joint project of the University of California, Los Angeles, and the Jet Propulsion 
Laboratory/California Institute of Technology, funded by the National Aeronautics and Space 
Administration.

The 2dF QSO Redshift Survey (2QZ) was compiled by the 2QZ survey team from observations made 
with the 2-degree Field on the Anglo-Australian Telescope.

\software{Numpy, Scipy, astropy, specDB, linetools, Pandas, SExtratcor, cmb\_footprint, Topcat}

\bibliographystyle{yahapj}
\bibliography{references}

\end{document}